\input hyperbasics

\input epsf
\catcode`\@=11
\def\unredoffs{\voffset=13mm \hoffset=6.5truemm}
\def\redoffs{\voffset=-12.truemm\hoffset=-3truemm}
\def\speclscape{}
\newbox\leftpage \newdimen\fullhsize \newdimen\hstitle \newdimen\hsbody
\newdimen\hdim
\hfuzz=1pt
\ifx\hyperdef\UNd@FiNeD\def\hyperdef#1#2#3#4{#4}\def\hyperref#1#2#3#4{#4}\fi
\def\newans{y }
\def\answb{y }
\ifx\answb\newans\message{(This uses normal fonts.)}%
\def\bigans{b }
\def\answ{b }
\ifx\answ\bigans\message{(Format simple colonne 12pts.}
\magnification=1200 \unredoffs\hsize=147truemm\vsize=219truemm
\hsbody=\hsize \hstitle=\hsize 
\else\message{(Format double colonne, 10pts.} \let\l@r=L
\magnification=1000 \vsize=182.5truemm
\redoffs%
\hstitle=122.5truemm\hsbody=122.5truemm\fullhsize=258truemm\hsize=\hsbody
\output={
  \almostshipout{\leftline{\vbox{\makeheadline\pagebody\makefootline}}}
\advancepageno%
}
\def\almostshipout#1{\if L\l@r \count1=1 \message{[\the\count0.\the\count1]}
      \global\setbox\leftpage=#1 \global\let\l@r=R
 \else \count1=2
  \shipout\vbox{\speclscape{\hsize\fullhsize}
      \hbox to\fullhsize{\box\leftpage\hfil#1}}  \global\let\l@r=L\fi}
\fi

\def\sla#1{\mkern-1.5mu\raise0.4pt\hbox{$\not$}\mkern1.2mu #1\mkern 0.7mu}
\def\Dbar{\mkern-1.5mu\raise0.4pt\hbox{$\not$}\mkern-.1mu {\rm D}\mkern.1mu}
\def\Abar{\mkern1.mu\raise0.4pt\hbox{$\not$}\mkern-1.3mu A\mkern.1mu}
\def\Bbar{\mkern-0.mu\raise0.4pt\hbox{$\not$}\mkern-.3mu B\mkern 0.6mu}
\newskip\tableskipamount \tableskipamount=8pt plus 3pt minus 3pt

  
\newdimen\chapskip
 \font\ssbx=cmssbx10
 \font\caprm=cmr9 \font\capit=cmti9
\font\capbf=cmbx9 \font\capsl=cmsl9 \font\capmi=cmmi9
\font\capex=cmex9 \font\capsy=cmsy9 \chapskip=17.5mm
\def\makeheadline{\vbox to 0pt{\vskip-22.5pt
\line{\vbox to8.5pt{}\the\headline}\vss}\nointerlineskip}
\font\tenbi=cmmib10 \font\ninebi=cmmib9 \font\sevenbi=cmmib7
\font\fivebi=cmmib5 \textfont4=\tenbi \scriptfont4=\sevenbi
\scriptscriptfont4=\fivebi \font\headrm=cmr10 

\font\sixrm=cmr6 \font\fiverm=cmr5 \font\sixmi=cmmi6
\font\fivemi=cmmi5 \font\sixsy=cmsy6 \font\fivesy=cmsy5
\font\sixbf=cmbx6 \font\fivebf=cmbx5 \skewchar\capmi='177
\skewchar\sixmi='177 \skewchar\fivemi='177 \skewchar\capsy='60
\skewchar\sixsy='60 \skewchar\fivesy='60

\def\elevenpoint{
\textfont0=\caprm \scriptfont0=\sixrm \scriptscriptfont0=\fiverm
\def\rm{\fam0\caprm}
\textfont1=\capmi \scriptfont1=\sixmi \scriptscriptfont1=\fivemi
\textfont2=\capsy \scriptfont2=\sixsy \scriptscriptfont2=\fivesy
\textfont3=\capex \scriptfont3=\capex \scriptscriptfont3=\capex
\textfont\itfam=\capit \def\it{\fam\itfam\capit} 
\textfont\slfam=\capsl  \def\sl{\fam\slfam\capsl} 
\textfont\bffam=\capbf \scriptfont\bffam=\sixbf
\scriptscriptfont\bffam=\fivebf
\def\bf{\fam\bffam\capbf} 
\textfont4=\ninebi \scriptfont4=\sevenbi
\scriptscriptfont4=\fivebi \abovedisplayskip=11pt plus 3pt minus
8pt \belowdisplayskip=\abovedisplayskip
\smallskipamount=2.7pt plus 1pt minus 1pt
\medskipamount=5.4pt plus 2pt minus 2pt
\bigskipamount=10.8pt plus 3.6pt minus 3.6pt
\normalbaselineskip=11pt \setbox\strutbox=\hbox{\vrule height7.8pt
depth3.2pt width0pt} \normalbaselines \rm}


\catcode`\@=11

\font\tenmsa=msam10 \font\sevenmsa=msam7 \font\fivemsa=msam5
\font\tenmsb=msbm10 \font\sevenmsb=msbm7 \font\fivemsb=msbm5
\newfam\msafam
\newfam\msbfam
\textfont\msafam=\tenmsa  \scriptfont\msafam=\sevenmsa
  \scriptscriptfont\msafam=\fivemsa
\textfont\msbfam=\tenmsb  \scriptfont\msbfam=\sevenmsb
  \scriptscriptfont\msbfam=\fivemsb

\def\hexnumber@#1{\ifcase#1 0\or1\or2\or3\or4\or5\or6\or7\or8\or9\or
    A\or B\or C\or D\or E\or F\fi }

\font\teneuf=eufm10 \font\seveneuf=eufm7 \font\fiveeuf=eufm5
\newfam\euffam
\textfont\euffam=\teneuf \scriptfont\euffam=\seveneuf
\scriptscriptfont\euffam=\fiveeuf
\def\frak{\ifmmode\let\next\frak@\else
 \def\next{\Err@{Use \string\frak\space only in math mode}}\fi\next}
\def\goth{\ifmmode\let\next\frak@\else
 \def\next{\Err@{Use \string\goth\space only in math mode}}\fi\next}
\def\frak@#1{{\frak@@{#1}}}
\def\frak@@#1{\fam\euffam#1}

\edef\msa@{\hexnumber@\msafam} \edef\msb@{\hexnumber@\msbfam}

\mathchardef\boxdot="2\msa@00 \mathchardef\boxplus="2\msa@01
\mathchardef\boxtimes="2\msa@02 \mathchardef\square="0\msa@03
\mathchardef\blacksquare="0\msa@04
\mathchardef\centerdot="2\msa@05 \mathchardef\lozenge="0\msa@06
\mathchardef\blacklozenge="0\msa@07
\mathchardef\circlearrowright="3\msa@08
\mathchardef\circlearrowleft="3\msa@09
\mathchardef\rightleftharpoons="3\msa@0A
\mathchardef\leftrightharpoons="3\msa@0B
\mathchardef\boxminus="2\msa@0C \mathchardef\Vdash="3\msa@0D
\mathchardef\Vvdash="3\msa@0E \mathchardef\vDash="3\msa@0F
\mathchardef\twoheadrightarrow="3\msa@10
\mathchardef\twoheadleftarrow="3\msa@11
\mathchardef\leftleftarrows="3\msa@12
\mathchardef\rightrightarrows="3\msa@13
\mathchardef\upuparrows="3\msa@14
\mathchardef\downdownarrows="3\msa@15
\mathchardef\upharpoonright="3\msa@16

\mathchardef\downharpoonright="3\msa@17
\mathchardef\upharpoonleft="3\msa@18
\mathchardef\downharpoonleft="3\msa@19
\mathchardef\rightarrowtail="3\msa@1A
\mathchardef\leftarrowtail="3\msa@1B
\mathchardef\leftrightarrows="3\msa@1C
\mathchardef\rightleftarrows="3\msa@1D \mathchardef\Lsh="3\msa@1E
\mathchardef\Rsh="3\msa@1F \mathchardef\rightsquigarrow="3\msa@20
\mathchardef\leftrightsquigarrow="3\msa@21
\mathchardef\looparrowleft="3\msa@22
\mathchardef\looparrowright="3\msa@23
\mathchardef\circeq="3\msa@24 \mathchardef\succsim="3\msa@25
\mathchardef\gtrsim="3\msa@26 \mathchardef\gtrapprox="3\msa@27
\mathchardef\multimap="3\msa@28 \mathchardef\therefore="3\msa@29
\mathchardef\because="3\msa@2A \mathchardef\doteqdot="3\msa@2B

\mathchardef\triangleq="3\msa@2C \mathchardef\precsim="3\msa@2D
\mathchardef\lesssim="3\msa@2E \mathchardef\lessapprox="3\msa@2F
\mathchardef\eqslantless="3\msa@30
\mathchardef\eqslantgtr="3\msa@31
\mathchardef\curlyeqprec="3\msa@32
\mathchardef\curlyeqsucc="3\msa@33
\mathchardef\preccurlyeq="3\msa@34 \mathchardef\leqq="3\msa@35
\mathchardef\leqslant="3\msa@36 \mathchardef\lessgtr="3\msa@37
\mathchardef\backprime="0\msa@38
\mathchardef\risingdotseq="3\msa@3A
\mathchardef\fallingdotseq="3\msa@3B
\mathchardef\succcurlyeq="3\msa@3C \mathchardef\geqq="3\msa@3D
\mathchardef\geqslant="3\msa@3E \mathchardef\gtrless="3\msa@3F
\mathchardef\sqsubset="3\msa@40 \mathchardef\sqsupset="3\msa@41
\mathchardef\vartriangleright="3\msa@42
\mathchardef\vartriangleleft="3\msa@43
\mathchardef\trianglerighteq="3\msa@44
\mathchardef\trianglelefteq="3\msa@45
\mathchardef\bigstar="0\msa@46 \mathchardef\between="3\msa@47
\mathchardef\blacktriangledown="0\msa@48
\mathchardef\blacktriangleright="3\msa@49
\mathchardef\blacktriangleleft="3\msa@4A
\mathchardef\vartriangle="0\msa@4D
\mathchardef\blacktriangle="0\msa@4E
\mathchardef\triangledown="0\msa@4F \mathchardef\eqcirc="3\msa@50
\mathchardef\lesseqgtr="3\msa@51 \mathchardef\gtreqless="3\msa@52
\mathchardef\lesseqqgtr="3\msa@53
\mathchardef\gtreqqless="3\msa@54
\mathchardef\Rrightarrow="3\msa@56
\mathchardef\Lleftarrow="3\msa@57 \mathchardef\veebar="2\msa@59
\mathchardef\barwedge="2\msa@5A
\mathchardef\doublebarwedge="2\msa@5B \mathchardef\angle="0\msa@5C
\mathchardef\measuredangle="0\msa@5D
\mathchardef\sphericalangle="0\msa@5E
\mathchardef\varpropto="3\msa@5F \mathchardef\smallsmile="3\msa@60
\mathchardef\smallfrown="3\msa@61 \mathchardef\Subset="3\msa@62
\mathchardef\Supset="3\msa@63 \mathchardef\Cup="2\msa@64

\mathchardef\Cap="2\msa@65

\mathchardef\curlywedge="2\msa@66 \mathchardef\curlyvee="2\msa@67
\mathchardef\leftthreetimes="2\msa@68
\mathchardef\rightthreetimes="2\msa@69
\mathchardef\subseteqq="3\msa@6A \mathchardef\supseteqq="3\msa@6B
\mathchardef\bumpeq="3\msa@6C \mathchardef\Bumpeq="3\msa@6D
\mathchardef\lll="3\msa@6E

\mathchardef\ggg="3\msa@6F

\mathchardef\circledS="0\msa@73 \mathchardef\pitchfork="3\msa@74
\mathchardef\dotplus="2\msa@75 \mathchardef\backsim="3\msa@76
\mathchardef\backsimeq="3\msa@77 \mathchardef\complement="0\msa@7B
\mathchardef\intercal="2\msa@7C \mathchardef\circledcirc="2\msa@7D
\mathchardef\circledast="2\msa@7E
\mathchardef\circleddash="2\msa@7F
\def\ulcorner{\delimiter"4\msa@70\msa@70 }
\def\urcorner{\delimiter"5\msa@71\msa@71 }
\def\llcorner{\delimiter"4\msa@78\msa@78 }
\def\lrcorner{\delimiter"5\msa@79\msa@79 }
\def\yen{\mathhexbox\msa@55 }
\def\checkmark{\mathhexbox\msa@58 }
\def\circledR{\mathhexbox\msa@72 }
\def\maltese{\mathhexbox\msa@7A }
\mathchardef\lvertneqq="3\msb@00 \mathchardef\gvertneqq="3\msb@01
\mathchardef\nleq="3\msb@02 \mathchardef\ngeq="3\msb@03
\mathchardef\nless="3\msb@04 \mathchardef\ngtr="3\msb@05
\mathchardef\nprec="3\msb@06 \mathchardef\nsucc="3\msb@07
\mathchardef\lneqq="3\msb@08 \mathchardef\gneqq="3\msb@09
\mathchardef\nleqslant="3\msb@0A \mathchardef\ngeqslant="3\msb@0B
\mathchardef\lneq="3\msb@0C \mathchardef\gneq="3\msb@0D
\mathchardef\npreceq="3\msb@0E \mathchardef\nsucceq="3\msb@0F
\mathchardef\precnsim="3\msb@10 \mathchardef\succnsim="3\msb@11
\mathchardef\lnsim="3\msb@12 \mathchardef\gnsim="3\msb@13
\mathchardef\nleqq="3\msb@14 \mathchardef\ngeqq="3\msb@15
\mathchardef\precneqq="3\msb@16 \mathchardef\succneqq="3\msb@17
\mathchardef\precnapprox="3\msb@18
\mathchardef\succnapprox="3\msb@19 \mathchardef\lnapprox="3\msb@1A
\mathchardef\gnapprox="3\msb@1B \mathchardef\nsim="3\msb@1C
\mathchardef\ncong="3\msb@1D

\mathchardef\varsubsetneq="3\msb@20
\mathchardef\varsupsetneq="3\msb@21
\mathchardef\nsubseteqq="3\msb@22
\mathchardef\nsupseteqq="3\msb@23
\mathchardef\subsetneqq="3\msb@24
\mathchardef\supsetneqq="3\msb@25
\mathchardef\varsubsetneqq="3\msb@26
\mathchardef\varsupsetneqq="3\msb@27
\mathchardef\subsetneq="3\msb@28 \mathchardef\supsetneq="3\msb@29
\mathchardef\nsubseteq="3\msb@2A \mathchardef\nsupseteq="3\msb@2B
\mathchardef\nparallel="3\msb@2C \mathchardef\nmid="3\msb@2D
\mathchardef\nshortmid="3\msb@2E
\mathchardef\nshortparallel="3\msb@2F
\mathchardef\nvdash="3\msb@30 \mathchardef\nVdash="3\msb@31
\mathchardef\nvDash="3\msb@32 \mathchardef\nVDash="3\msb@33
\mathchardef\ntrianglerighteq="3\msb@34
\mathchardef\ntrianglelefteq="3\msb@35
\mathchardef\ntriangleleft="3\msb@36
\mathchardef\ntriangleright="3\msb@37
\mathchardef\nleftarrow="3\msb@38
\mathchardef\nrightarrow="3\msb@39
\mathchardef\nLeftarrow="3\msb@3A
\mathchardef\nRightarrow="3\msb@3B
\mathchardef\nLeftrightarrow="3\msb@3C
\mathchardef\nleftrightarrow="3\msb@3D
\mathchardef\divideontimes="2\msb@3E
\mathchardef\varnothing="0\msb@3F \mathchardef\nexists="0\msb@40
\mathchardef\mho="0\msb@66 \mathchardef\eth="0\msb@67
\mathchardef\eqsim="3\msb@68 \mathchardef\beth="0\msb@69
\mathchardef\gimel="0\msb@6A \mathchardef\daleth="0\msb@6B
\mathchardef\lessdot="3\msb@6C \mathchardef\gtrdot="3\msb@6D
\mathchardef\ltimes="2\msb@6E \mathchardef\rtimes="2\msb@6F
\mathchardef\shortmid="3\msb@70
\mathchardef\shortparallel="3\msb@71
\mathchardef\smallsetminus="2\msb@72
\mathchardef\thicksim="3\msb@73 \mathchardef\thickapprox="3\msb@74
\mathchardef\approxeq="3\msb@75 \mathchardef\succapprox="3\msb@76
\mathchardef\precapprox="3\msb@77
\mathchardef\curvearrowleft="3\msb@78
\mathchardef\curvearrowright="3\msb@79
\mathchardef\digamma="0\msb@7A \mathchardef\varkappa="0\msb@7B
\mathchardef\hslash="0\msb@7D \mathchardef\hbar="0\msb@7E
\mathchardef\backepsilon="3\msb@7F
\def\Bbb{\ifmmode\let\next\Bbb@\else
 \def\next{\errmessage{Use \string\Bbb\space only in math mode}}\fi\next}
\def\Bbb@#1{{\Bbb@@{#1}}}
\def\Bbb@@#1{\fam\msbfam#1}
 \catcode`\@=12
\def\sla#1{\mkern-1.5mu\raise0.4pt\hbox{$\not$}\mkern1.2mu #1\mkern 0.7mu}
\def\Dbar{\mkern-1.5mu\raise0.4pt\hbox{$\not$}\mkern-.1mu {\rm D}\mkern.1mu}
\def\Abar{\mkern1.mu\raise0.4pt\hbox{$\not$}\mkern-1.3mu A\mkern.1mu}
\nopagenumbers
\headline={\ifnum\pageno=1\hfill\else\draftdate\hfil{\headrm\folio}%
\hfil\hphantom{\draftdate}\fi} \else\message{(This uses pseudo
12pts fonts.} \hoffset=8mm \voffset=16mm
\input lfont12 

\def\sla#1{\mkern-1.5mu\raise0.5pt\hbox{$\not$}\mkern1.2mu #1\mkern 0.7mu}
\def\Dbar{\mkern-1.5mu\raise0.5pt\hbox{$\not$}\mkern-.1mu {\rm D}\mkern.1mu}
\def\Abar{\mkern1.mu\raise0.5pt\hbox{$\not$}\mkern-1.3mu A\mkern.1mu}
\fi


\newcount\yearltd\yearltd=\year\advance\yearltd by -2000
\newif\ifdraftmode
\draftmodefalse
\def\draft{\draftmodetrue{\count255=\time\divide\count255 by 60
\xdef\hourmin{\number\count255}
  \multiply\count255 by-60\advance\count255 by\time
  \xdef\hourmin{\hourmin:\ifnum\count255<10 0\fi\the\count255}}}
\def\draftdate{\ifdraftmode{\headrm\quad (\jobname,\ le
\number\day/\number\month/\number\yearltd\ \ \hourmin)}\else{}\fi}
\newif\iffrancmode
\francmodefalse
\def\e{\mathop{\rm e}\nolimits}

\def\d{{\rm d}}
\def\ud{{\textstyle{1\over 2}}}
\def\half{\ud}
\def\tr{\mathop{\rm tr}\nolimits}

\def\del{\partial}

\chardef\sigmat=27

\def\frac#1#2{{\textstyle{#1\over#2}}}

\def\leaderfill{\leaders\hbox to 1em{\hss.\hss}\hfill}
\catcode`\@=11
\def\deqalignno#1{\displ@y\tabskip\centering \halign to
\displaywidth{\hfil$\displaystyle{##}$\tabskip0pt&$\displaystyle{{}##}$
\hfil\tabskip0pt &\quad
\hfil$\displaystyle{##}$\tabskip0pt&$\displaystyle{{}##}$
\hfil\tabskip\centering& \llap{$##$}\tabskip0pt \crcr #1 \crcr}}
\def\deqalign#1{\null\,\vcenter{\openup\jot\m@th\ialign{
\strut\hfil$\displaystyle{##}$&$\displaystyle{{}##}$\hfil
&&\quad\strut\hfil$\displaystyle{##}$&$\displaystyle{{}##}$
\hfil\crcr#1\crcr}}\,}
\def\xlabel#1{\expandafter\xl@bel#1}\def\xl@bel#1{#1}
\def\label#1{\l@bel #1{\hbox{}}}
\def\l@bel#1{\ifx\UNd@FiNeD#1\message{label \string#1 is undefined.}%
\xdef#1{?.? }\fi{\let\hyperref=\relax\xdef\next{#1}}%
\ifx\next\em@rk\def\next{}%
\else\def\next{#1}\fi\next}
\def\DefWarn#1{\ifx\UNd@FiNeD#1\else
\immediate\write16{*** WARNING: the label \string#1 is already defined%
***}\fi}%
\newread\ch@ckfile
\def\cinput#1{\def\filen@me{#1 }
\immediate\openin\ch@ckfile=\filen@me
\ifeof\ch@ckfile\message{<< (\filen@me\ DOES NOT EXIST in this pass)>>}\else%
\closein \ch@ckfile\input\filen@me\fi}
\ifx\UNd@FiNeD\prefix\def\prefix{}\fi 
\newread\ch@ckfile
\immediate\openin\ch@ckfile=\jobname.def
\ifeof\ch@ckfile\message{<< (\jobname.def DOES NOT EXIST in this
pass) >>} \else
\def\DefWarn#1{}%
\closein \ch@ckfile%
\input\jobname.def\fi
\def\listcontent{
\immediate\openin\ch@ckfile=\jobname.tab 
\ifeof\ch@ckfile\message{no file \jobname.tab, no table of
contents this
pass}%
\else\closein\ch@ckfile\centerline{\bf\iffrancmode Table des
mati\`eres \else Contents\fi}\nobreak\medskip%
{\baselineskip=12pt\parskip=0pt\catcode`\@=11\input\jobname.tab
\catcode`\@=12\bigbreak\bigskip}\fi}
\newcount\nosection
\newcount\nosubsection
\newcount\neqno
\newcount\notenumber
\newcount\nofigure
\newcount\notable
\newcount\noexerc
\newif\ifappmode
\def\equation{\jobname.equ}
\newwrite\equa

\newdimen\hulp
\def\maketitle#1{
\edef\oneliner##1{\centerline{##1}}
\edef\twoliner##1{\vbox{\parindent=0pt\leftskip=0pt plus
1fill\rightskip=0pt plus 1fill
                     \parfillskip=0pt\relax##1}}
\setbox0=\vbox{#1}\hulp=0.5\hsize
                 \ifdim\wd0<\hulp\oneliner{#1}\else
                 \twoliner{#1}\fi}
\def\preprint#1{\ifdraftmode\gdef\prepname{\jobname/#1}\else%
\gdef\prepname{#1}\fi\hfill{
\expandafter{\prepname}}\vskip20mm}
\def\title#1\par{\gdef\titlename{#1}
\maketitle{\ssbx\uppercase\expandafter{\titlename}} \vskip20truemm
\nosection=0 \neqno=0 \notenumber=0 \nofigure=0 \notable=0
\def\prefix{}
\appmodefalse \mark{\the\nosection} \message{#1}
\immediate\openout\equa=\equation }
\def\abstract{\vskip8mm\iffrancmode\centerline{R\'ESUM\'E}\else%
\centerline{ABSTRACT}\fi \smallskip \begingroup\narrower
\elevenpoint\baselineskip10pt}
\def\endabstract{\par\endgroup \bigskip}
\def\section#1\par{\vskip0pt plus.1\vsize\penalty-100\vskip0pt plus-.1
\vsize\bigskip\vskip\parskip
\ifnum\nosection=0\ifappmode\relax\else\writetoc
\fi\fi
\advance\nosection by 1\global\nosubsection=0\global\neqno=0
\vbox{\noindent\bf{\hyperdef\hypernoname{section}{\prefix\the\nosection}%
{\prefix\the\nosection}\ #1}}
\writetoca{{\string\hyperref{}{section}{\prefix\the\nosection}%
{\prefix\the\nosection}} {#1}} \message{\the\nosection\ #1}
\mark{\the\nosection}\bigskip\noindent }
\def\appendix#1#2\par{\bigbreak \nosection=0 \appmodetrue \notenumber=0 \neqno=0
\def\prefix{A}
\mark{\the\nosection} \message{APPENDICES}
{\centerline{\bf Appendices}
\hyperdef\hypernoname{appendix}{\prefix}{
\leftline{\uppercase\expandafter{#1}}
\leftline{\uppercase\expandafter{#2}}}}
\writetoca{\string\hyperref{}{appendix}{\prefix}{Appendices}.\ #1 \ #2}%
}
\def\subsection#1\par {\vskip0pt plus.05\vsize\penalty-100\vskip0pt
plus-.05\vsize\bigskip\vskip\parskip\advance\nosubsection by 1
\vbox{\noindent\it{\hyperdef\hypernoname{subsection}{\prefix\the\nosection.%
\the\nosubsection}{\prefix\the\nosection.\the\nosubsection\ #1}}}%
\smallskip\noindent
\writetoca{{\string\hyperref{}{subsection}{\prefix\the\nosection.%
\the\nosubsection}{\prefix\the\nosection.\the\nosubsection}} {#1}}
\message{\the\nosection.\the\nosubsection\ #1} }
\def\note #1{\advance\notenumber by 1
\footnote{$^{\the\notenumber}$}{\sevenrm #1}}

\parindent=1em
\newinsert\margin
\dimen\margin=\maxdimen \count\margin=0 \skip\margin=0pt
\def\sslbl#1{\DefWarn#1%
\ifdraftmode{\hfill\escapechar-1{\rlap{\hskip-1mm%
\sevenrm\string#1}}}\fi%
\ifnum\nosection=0\if\prefix{}\xdef#1{}%
\edef\ewrite{\write\equa{{\string#1}}%
\write\equa{}}\ewrite%
\else
\xdef#1{\noexpand\hyperref{}{appendix}{\prefix}{\prefix}}%
\edef\ewrite{\write\equa{{\string#1},\prefix}%
\write\equa{}}\ewrite%
\writedef{#1\leftbracket#1} \fi
\else%
\ifnum\nosubsection=0%
\xdef#1{\noexpand\hyperref{}{section}{\prefix\the\nosection}%
{\prefix\the\nosection}}%
\edef\ewrite{\write\equa{{\string#1},\prefix\the\nosection}%
\write\equa{}}\ewrite%
\writedef{#1\leftbracket#1}
\else%
\xdef#1{\noexpand\hyperref{}{subsection}{\prefix\the\nosection.%
\the\nosubsection}{\prefix\the\nosection.\the\nosubsection}}%
\writedef{#1\leftbracket#1}
\edef\ewrite{\write\equa{{\string#1},\prefix\the\nosection%
.\the\nosubsection}\write\equa{}}\ewrite\fi\fi}%
\newwrite\tfile \def\writetoca#1{}
\def\writetoc{\immediate\openout\tfile=\jobname.tab
\def\writetoca##1{{\edef\next{\write\tfile{\noindent ##1 \string\leaderfill%
\noexpand\number\pageno\par}}\next}}}

%
\def\nolabels{\def\wrlabeL##1{}\def\eqlabeL##1{}\def\reflabeL##1{}}
\def\writelabels{\def\wrlabeL##1{\leavevmode\vadjust{\rlap{\smash%
{\line{{\escapechar=` \hfill\rlap{\sevenrm\hskip.03in\string##1}}}}}}}%
\def\eqlabeL##1{{\escapechar-1\rlap{\sevenrm\hskip.05in\string##1}}}%
\def\reflabeL##1{\noexpand\llap{\noexpand\sevenrm\string\string\string##1}}}
\nolabels

\global\newcount\refno \global\refno=1
\newwrite\rfile
\def\ref{[\hyperref{}{reference}{\the\refno}{\the\refno}]\nref}
\def\nref#1{\DefWarn#1%
\xdef#1{[\noexpand\hyperref{}{reference}{\the\refno}{\the\refno}]}%
\writedef{#1\leftbracket#1}%
\ifnum\refno=1\immediate\openout\rfile=\jobname.ref\fi
\chardef\wfile=\rfile\immediate\write\rfile{\noexpand\item{[\noexpand\hyperdef%
\noexpand\hypernoname{reference}{\the\refno}{\the\refno}]\ }%
\reflabeL{#1\hskip.31in}\pctsign}\global\advance\refno
by1\findarg}
\def\findarg#1#{\begingroup\obeylines\newlinechar=`\^^M\pass@rg}
{\obeylines\gdef\pass@rg#1{\writ@line\relax #1^^M\hbox{}^^M}%
\gdef\writ@line#1^^M{\expandafter\toks0\expandafter{\striprel@x #1}%
\edef\next{\the\toks0}\ifx\next\em@rk\let\next=\endgroup\else\ifx\next\empty%
\else\immediate\write\wfile{\the\toks0}\fi\let\next=\writ@line\fi\next\relax}}
\def\striprel@x#1{} \def\em@rk{\hbox{}}
\def\lref{\begingroup\obeylines\lr@f}
\def\lr@f#1#2{\DefWarn#1\gdef#1{\let#1=\UNd@FiNeD\ref#1{#2}}\endgroup\unskip}

\def\addref#1{\immediate\write\rfile{\noexpand\item{}#1}} 
\def\listrefs{{}\vfill\supereject\immediate\closeout\rfile\writestoppt
\baselineskip=14pt\centerline{{\bf\iffrancmode R\'eferences\else References%
\fi}}
\bigskip{\parindent=20pt%
\frenchspacing\escapechar=` \input
\jobname.ref\vfill\eject}\nonfrenchspacing}
\def\startrefs#1{\immediate\openout\rfile=\jobname.ref\refno=#1}
\def\xref{\expandafter\xr@f}\def\xr@f[#1]{#1}
\def\refs#1{\count255=1[\r@fs #1{\hbox{}}]}
\def\r@fs#1{\ifx\UNd@FiNeD#1\message{reflabel \string#1 is undefined.}%
\nref#1{need to supply reference \string#1.}\fi%
\vphantom{\hphantom{#1}}{\let\hyperref=\relax\xdef\next{#1}}%
\ifx\next\em@rk\def\next{}%
\else\ifx\next#1\ifodd\count255\relax\xref#1\count255=0\fi%
\else#1\count255=1\fi\let\next=\r@fs\fi\next}
\newwrite\lfile
{\escapechar-1\xdef\pctsign{\string\%}\xdef\leftbracket{\string\{}
\xdef\rightbracket{\string\}}\xdef\numbersign{\string\#}}
\def\writedefs{\immediate\openout\lfile=\jobname.def \def\writedef##1{%
{\let\hyperref=\relax\let\hyperdef=\relax\let\hypernoname=\relax
 \immediate\write\lfile{\string\def\string##1\rightbracket}}}}%
\def\writestop{\def\writestoppt{\immediate\write\lfile{\string\pageno%
\the\pageno\string\startrefs\leftbracket\the\refno\rightbracket%
\string\def\string\secsym\leftbracket\secsym\rightbracket%
\string\secno\the\secno\string\meqno\the\meqno}\immediate\closeout\lfile}}
\def\writestoppt{}\def\writedef#1{}
\writedefs
\def\biblio\par{\vskip0pt plus.1\vsize\penalty-100\vskip0pt plus-.1
\vsize\bigskip\vskip\parskip \message{Bibliographie}
{\leftline{\bf \hyperdef\hypernoname{biblio}{bib}{Bibliographical
Notes}}} \nobreak\medskip\noindent\frenchspacing
\writetoca{\string\hyperref{}{biblio}{bib}{Bibliographical Notes}}}%

\def\biblionote{\iffrancmode Notes Bibliographiques\else Bibliographical Notes
\fi}
\def\beginbib\par{\vskip0pt plus.1\vsize\penalty-100\vskip0pt plus-.1
\vsize\bigskip\vskip\parskip \message{Bibliographie}
{\leftline{\bf \hyperdef\hypernoname{biblio}{\the\nosection}%
{\biblionote}}} \nobreak\medskip\noindent\frenchspacing
\writetoca{\string\hyperref{}{biblio}{\the\nosection}%
{\biblionote}}}%

\def\Exercises{\iffrancmode Exercices\else Exercises
\fi}
\def\exerc\par{\vskip0pt plus.1\vsize\penalty-100\vskip0pt plus-.1
\vsize\bigskip\vskip\parskip\global\noexerc=0 \message{Exercises}
{\leftline{\bf
\hyperdef\hypernoname{exercise}{\the\nosection}{\Exercises}}}
\nobreak\medskip\noindent\frenchspacing
\writetoca{\string\hyperref{}{exercise}{\the\nosection}{\Exercises}}
}
\def\esubsec{\ifnum\noexerc=0\vskip-12pt\else\vskip0pt plus.05\vsize%
\penalty-100\vskip0pt plus-.05\vsize\bigskip\vskip\parskip\fi%
\global\advance\noexerc by 1
\hyperdef\hypernoname{exercise}{\the\nosection.\the\noexerc}{}%
\vbox{\noindent\it \iffrancmode Exercice\else Exercise\fi\
\the\nosection.\the\noexerc}\smallskip\noindent}
\def\exelbl#1{\ifdraftmode{\hfill\escapechar-1{\rlap{\hskip-1mm%
\sevenrm\string#1}}}\fi%
{\xdef#1{\noexpand\hyperref{}{exercise}{\the\nosection.\the\noexerc}%
{\the\nosection.\the\noexerc}}}%
\edef\ewrite{\write\equa{{\string#1}\the\nosection.\the\noexerc}%
\write\equa{}}\ewrite%
\writedef{#1\leftbracket#1}}
\def\eqnn{\global\advance\neqno by 1 \ifinner\relax\else%
\eqno\fi(\prefix\the\nosection.\the\neqno)}
%
\def\eqnd#1{\DefWarn#1%
\global\advance\neqno by 1
{\xdef#1{($\noexpand\hyperref{}{equation}{\prefix\the\nosection.\the\neqno}%
{\prefix\the\nosection.\the\neqno}$)}}
\ifinner\relax\else\eqno\fi(\hyperdef\hypernoname{equation}{\prefix\the%
\nosection.\the\neqno}{\prefix\the\nosection.\the\neqno})
\writedef{#1\leftbracket#1}
\ifdraftmode{\escapechar-1{\rlap{\hskip.2mm\sevenrm\string#1}}}\fi
\edef\ewrite{\write\equa{{\string#1},(\prefix\the\nosection.\the\neqno)
{\noexpand\number\pageno}}\write\equa{}}\ewrite}
%
\def\checkm@de#1#2{\ifmmode{\def\f@rst##1{##1}\hyperdef\hypernoname{equation}%
{#1}{#2}}\else\hyperref{}{equation}{#1}{#2}\fi}
\def\f@rst#1{\c@t#1a\em@ark}\def\c@t#1#2\em@ark{#1}
\def\eqna#1{\DefWarn#1%
\global\advance\neqno by1\ifdraftmode{\hfill%
\escapechar-1{\rlap{\sevenrm\string#1}}}\fi%
\xdef #1##1{(\noexpand\relax\noexpand%
\checkm@de{\prefix\the\nosection.\the\neqno\noexpand\f@rst{##1}1}%
{\hbox{$\prefix\the\nosection.\the\neqno##1$}})}
\writedef{#1\numbersign1\leftbracket#1{\numbersign1}}%
}
%

%
\def\em@rk{\hbox{}}
\def\xeqn{\expandafter\xe@n}\def\xe@n(#1){#1}
\def\xeqna#1{\expandafter\xe@na#1}\def\xe@na\hbox#1{\xe@nap #1}
\def\xe@nap$(#1)${\hbox{$#1$}}
\def\eqns#1{(\e@ns #1{\hbox{}})}
\def\e@ns#1{\ifx\UNd@FiNeD#1\message{eqnlabel \string#1 is undefined.}%
\xdef#1{(?.?)}\fi{\let\hyperref=\relax\xdef\next{#1}}%
\ifx\next\em@rk\def\next{}%
\else\ifx\next#1\xeqn#1\else\def\n@xt{#1}\ifx\n@xt\next#1\else\xeqna#1\fi
\fi\let\next=\e@ns\fi\next}
\def\figure#1#2{\global\advance\nofigure by 1 \vglue#1%
\hyperdef\hypernoname{figure}{\the\nofigure}{}%
{\elevenpoint \setbox1=\hbox{#2}
\ifdim\wd1=0pt\centerline{Fig.\ \the\nofigure\hskip0.5mm}%
\else\def\caption{Fig.\ \the\nofigure\quad#2\hskip0mm}
\setbox0=\hbox{\caption} \ifdim\wd0>\hsize\noindent Fig.\
\the\nofigure\quad#2\else
                 \centerline{\caption}\fi\fi}\par}
\def\lfigure#1#2{\global\advance\nofigure by
1\vglue#1%
\hyperdef\hypernoname{figure}{\the\nofigure}{}%
\leftline{\elevenpoint\hskip10truemm  Fig.\ \the\nofigure\quad
#2}}
\def\figlbl#1{\ifdraftmode{\hfill\escapechar-1{\rlap{\hskip-1mm%
\sevenrm\string#1}}}\fi%
{\xdef#1{\noexpand\hyperref{}{figure}{\the\nofigure}%
{\the\nofigure}}}%
\edef\ewrite{\write\equa{{\string#1}\the\nofigure}%
\write\equa{}}\ewrite%
\writedef{#1\leftbracket#1}}
\def\tablbl#1{\global\advance\notable by
1\ifdraftmode{\hfill\escapechar-1{\rlap{\hskip-1mm%
\sevenrm\string#1}}}\fi%
{\xdef#1{\noexpand\hyperref{}{table}{\the\notable}%
{\the\notable}}}%
\hyperdef\hypernoname{table}{\the\notable}{}%
\edef\ewrite{\write\equa{{\string#1}\the\notable}%
\write\equa{}}\ewrite%
\writedef{#1\leftbracket#1}}

\catcode`@=12




\francmodefalse

\def\*{********** new, please check **********}

\def\slam#1{\mkern-1.5mu\raise-0.2pt\hbox{$\scriptstyle{\not}$}\mkern1.2mu
#1\mkern 0.7mu}


\def\D{{\rm D}}


\def\half{\frac{1}{2}}
\def\lb{\hfil\break}
\def\del{\partial}

\def\delslash{\mathord{\not\mathrel{\partial}}}

\def\ma{m_\varphi}
\def\mpsi{m_\psi}
\def\del{\partial}

\def\sqr#1#2{{\vcenter{\hrule height.#2pt
     \hbox{\vrule width.#2pt height#1pt \kern#1pt
          \vrule width.#2pt}
       \hrule height.#2pt}}}

\def\to{\rightarrow}
\def\half{\frac{1}{2}}
\def\del{\partial}

\newskip\tableskipamount \tableskipamount=10pt plus 4pt minus 4pt

\def\upar{\uparrow\kern-9.\exec1pt\lower.2pt \hbox{$\uparrow$}}


\preprint{\lb Saclay-SPhT T02/110}
\hskip 4cm

\title{Phase Structure of Supersymmetric Models at Finite Temperature}

\centerline {   {\bf ~~Moshe~Moshe}${}^{a}$ ~~and~~ {\bf
~~Jean~Zinn-Justin}${}^{b}$}
\bigskip
{\baselineskip14pt
\smallskip
 \centerline{${}^a$ Department of
Physics, Technion -- Israel Institute of Technology,}
\centerline{Haifa, 32000 ISRAEL}
\centerline{${}^b$ CEA-Saclay, Service de Physique
Th\'eorique}
\centerline{ F-91191 Gif-sur-Yvette Cedex, FRANCE}
\centerline{and}
\centerline{Institut de Math\'ematiques  de Jussieu--Chevaleret,
Universit\'e de Paris VII}}

\footnote{}{(a)~Supported in part by the Israel Science Foundation
        grant number 193-00}

\footnote{}{~~~~~and the VPR fund for promotion of research.}


\footnote{}{(a)~email: moshe@physics.technion.ac.il}
\footnote{}{(b)~email: zinn@spht.saclay.cea.fr}

\vskip6mm


\abstract We study $O(N)$ symmetric supersymmetric models in three
dimensions at finite temperature. These models are known to have
an interesting phase structures. In particular, in the limit $N
\to \infty$ one finds spontaneous breaking of scale invariance
with no explicit breaking. Supersymmetry is softly broken at
finite temperature and the peculiar phase structure and properties
seen at $T=0$ are studied here at finite temperature.
\endabstract
\vfill\eject
\nref\Girard{L.Girardelo, M.T. Grisaru, P. Salomonson {\it Nucl.
Phys.} B178 (1981) 331 . }

\nref\Das{A.Das and M.Kaku {\it Phys. Rev} D18 (1978) 4540; A. Das
{\it Physics A } 158 (1989) 1, and references therein}

\nref\Senjan{ A. Riotto and G. Senjanovi\'c {\it Phys. Rev. Lett.} 79
(1997) 349 [arXiv:hep-ph/9702319]  and references therein }

\nref\Fotop{A. Fotopoulos and T.R. Taylor {\it Phys. Rev.} D59
(1999) 061701 and references therein }

\nref\Espin{J.R. Espinosa {\it Nucl. Phys.} B475 (1996) 273 ; M.
Laine {\it Nucl. Phys.} B481 (1996) 43 , Erratum {\it ibid.} B548
(1999) 637 }

\nref\BBM {W.A. Bardeen, M. Bander and M. Moshe, {\it Phys. Rev.
Lett.}  52  (1983) 1188. }

\nref\DKN{
 F.David, D.A. Kessler and H. Neuberger, {\it Phys. Rev. Lett.}
53  (1984) 2071 and {\it Nucl. Phys.}  B257 (1985) 695; D. A.
Kessler and H. Neuberger, {\it Phys. Lett.}
 157B  (1985) 416.}
\nref\Guds{ R. Gudmundsdottir, G. Rydnell and P. Salomonson, {\it
Phys. Rev. Lett. } 53 (1984) 2529;
Y. Matsubara, T. Suzuki and I. Yotsuyanagi, {\it Z. Phys.}
 C27  (1985) 599. } \vskip .2cm

\nref\BHM {W.A. Bardeen, K. Higashijima and M. Moshe, {\it Nucl.
Phys.}
 B250  (1985) 437, T.
Reisin M.Sc. Thesis -Technion  (March 1986)}

\section Introduction

Bosons and fermions behave differently as they interact with an
heat bath and supersymmetry is softly broken. Breakdown of
supersymmetry at finite temperature has attracted much attention
since the early interest in supersymmetry and involved much
controversy on its consequences, appearance and phase structure
\refs{\Girard, \Das}. Other related issues such as the restoration
of broken internal symmetries at finite temperature supersymmetric
theories were also debated \refs{\Senjan}. More recently, there is
a continued interest in the thermodynamics of supersymmetric
Yang-Mills theory \refs{\Fotop} and temperature effects on the
minimal supersymmetric model \refs{\Espin}.

Two supersymmetric models are studied in this paper at finite
temperature, the $O(N)$ symmetric supersymmetric $\vec\Phi^4$
field theory, which is shown to have a peculiar phase structure in
the large $N$ limit, and a supersymmetric non-linear
$\sigma$-model.

At zero temperature, it has been found in the past
\refs{\BBM,\DKN,\Guds} that the large $N$ limit of the scalar
$\vec \phi^6$ theory in $d=3$ dimensions and the supersymmetric
$\vec \Phi^4$ theory \refs{\BHM} in $d=3$ posses spontaneous
breaking of scale invariance unaccompanied by explicit breaking.
Here, we find out which of the peculiar properties of the phase
structure seen at $T=0$ in these models are maintained at finite
temperature and how phase transitions occur as the temperature,
rather than the coupling constant, is varied.

\section Supersymmetric models in the large $N$ limit

A simple supersymmetric model, with one scalar superfield, in
three euclidean dimensions at zero temperature is considered
first. We then examine a supersymmetric non-linear $\sigma$-model.
\subsection The supersymmetric scalar field in three dimensions

We consider the $O(N)$ invariant action
$${\cal S}(\Phi)=\int\d^3 x\,\d^2\theta\left[\ud\bar \D \Phi\cdot
\D \Phi+NU(\Phi^2/N)\right], \eqnd\eSUSYact $$ where $\Phi$ is an
$N$-component vector:
$~\Phi(\theta)=\varphi+\bar\theta\psi+\ud\bar\theta \theta F $.
Here, $\theta$ is a two-component Majorana spinor, $\varphi$ is an
$N$-component real scalar field, $\psi$ is an
\hbox{$N$-component,} two-component Majorana spinor  and $F~$ is
an $N$-component auxiliary field.  $\D =\partial /
\partial \bar \theta -\delslash ~\theta $  is the covariant derivative, the integration
measure is $\d^2\theta={i\over 2}d\theta_2 d\theta_1$ and
$\bar\theta_\alpha\theta_\beta =
\half\delta_{\alpha\beta}\bar\theta\theta $. More details about the
conventions used in the article can be found in the appendix.

In component form for a generic super-potential,  one finds
$$\eqalignno{ {\cal S}&=\int\d^3 x\,\left[-\ud \bar\psi\sla{\partial}\psi
+\ud (\partial_\mu\varphi)^2-\ud  U'(\varphi^2/N)\bar\psi\psi
-U''(\varphi^2/N)(\bar\psi\varphi)(\varphi\psi)/N
\right.\cr&\quad\left.+\ud \varphi^2
U'{}^2(\varphi^2/N)\right].&\eqnd\eGenericAction \cr}$$ The theory
violates parity symmetry. Space reflection is equivalent to the
change $U\mapsto -U$

One of the subjects which will be of interest here is spontaneous
breaking of scale invariance. When the binding potential
$U(\Phi^2)$ that binds bosonic and fermionic $O(N)$ quanta is
tuned to a particular strength at which $O(N)$ singlet massless
bound states are created, spontaneous breaking of scale invariance
occurs and the resulting massless Goldstone particles appear as a
supersymmetric multiplet of a dilaton and "dilatino". Though one
may expect this as a result of non-renormalization of the coupling
constant, in fact this happens also in the non-supersymmetric
case. Explicit breaking of scale invariance will appear only at
the next to leading order in $1/N$ as in the scalar case in the
$d=3$. At $ N \to \infty $ flat directions in parameter space
appear in the supersymmetric theory while maintaining the system's
ground state energy at zero.


\smallskip
\subsection Superfield formulation of large $N$ steepest descent
method.

In the large $N$ limit, a constraint on $\Phi^2$ is introduced by
adding to the initial action the term
$${\cal S}_L=\int\d^3x\,
\d^2\theta\,L(\theta)\left[\Phi^2(\theta)-NR(\theta)\right],
\eqnd\eactLagm $$ and integrating over  the superfields $
L(\theta)=M+\bar\theta\ell+\ud\bar\theta\theta \lambda$ and $
R(\theta)=\rho+\bar\theta\sigma+\ud\bar\theta\theta s$. $N-1$
components  of $\Phi$ are first integrated out while keeping a
test-component superfield $\Phi_1\equiv \phi$. One finds
$${\cal Z}=\int[\d\phi][\d R][\d L]\e^{-{\cal
S}_N(\phi,R,L)} \eqnd\eZSUSYfiv $$
with the large $N$ action
$$\eqalignno{{\cal S}_N&=\int\d^3 x\,\d^2\theta\left[\ud\bar \D \phi
\D \phi+NU(R)+ L \left(\phi^2-NR\right)\right] \cr
&\quad+\ud(N-1){\rm Str}\, \ln\left[-\bar \D \D
+2L\right].&\eqnd\eactSUSYN \cr}$$ The two saddle point equations,
obtained by varying the superfields $\phi$ and $R$, are
\eqna\eSUSYsad
$$\eqalignno{2L\phi-\bar \D  \D \phi&=0\,,&\eSUSYsad{a} \cr
L-U'(R)&=0\,.&\eSUSYsad{b} \cr} $$ The third saddle point
equation, obtained by varying $L$, involves the $\phi$
super-propagator given by the solution to the equation:
$$\left(- \bar \D  \D +2L(\theta)\right)\Delta(k,\theta,\theta')=
\delta^2(\theta'-\theta), $$
that is,
$$\Delta(k,\theta,\theta')={\left[1+\ud
M(\bar\theta\theta+\bar\theta'\theta') -\frac{1}{4}(\lambda+k^2)
\bar\theta\theta\bar\theta'\theta'\right]\over k^2+M^2+\lambda}
+{\bar\theta[i\sla{k}-M]\theta'\over k^2+M^2} \,.\eqnd\esupprop $$

At coinciding $\theta$ arguments, one finds
$$\Delta(k,\theta,\theta)={1+ M\bar\theta\theta\over
k^2+M^2+\lambda}-{M \bar\theta \theta\over k^2+M^2} \,.
\eqnd\eSUSYpropcoin $$ The trace over the group indices yields a
factor $(N-1)$. The third saddle point equation, obtained by
varying $L$ in Eq.~\eactSUSYN, is thus (for $N \gg 1$)
$$\eqalignno{R-\phi^2/N&={1\over N}\tr\Delta(k,\theta,\theta) \cr
&= {1\over (2\pi)^3} \int\d^3 k\left[{1+ M\bar\theta\theta \over
k^2+M^2+\lambda}-{M\bar\theta\theta\over k^2+M^2}\right].
&\eqnd\eSUSYsadc \cr}$$
\medskip
It is now convenient to introduce a notation for the boson mass:
$$m_\varphi\equiv m=\sqrt{M^2+\lambda }\,. \eqnn $$ and write
the saddle point equations in component form. Eq.~\eSUSYsad{a}
implies \eqna\eSUSYsada
$$\eqalignno{F-M\varphi&=0 \,, &\eSUSYsada {a}  \cr
\lambda\varphi+MF&=0\,. &  \eSUSYsada{b}      \cr}$$ Eliminating
$F$ between the two equations one finds $\varphi m^2=0 $ and,
thus, if the $O(N)$ symmetry is broken the boson mass $m_\varphi$
vanishes. Eq.~\eSUSYsad{b} yields \eqna\eSUSYsadb
$$\eqalignno{M&=U'(\rho),  & \eSUSYsadb{a} \cr
\lambda=m^2-M^2&=sU''(\rho).&\eSUSYsadb{b} \cr} $$

Eq.~\eSUSYsadc ~in component form is \eqna\eSUSYii
$$\eqalignno{\rho-\varphi^2/N&={1\over(2\pi)^3}\int{\d^3 p\over p^2+ m^2}
&\eSUSYii{a}\cr s-2F\varphi/N &={2M\over(2\pi)^3}\int\d^3
p\left({1\over p^2+m^2} -{1\over p^2+M^2}\right). &\eSUSYii{b}\cr}
$$

In these equations,  a cut-off $\Lambda$ is being introduced
explicitly.   We  set
$${1\over (2\pi)^d }\int^\Lambda {\d^d k \over k^2
+m^2}\equiv   \Omega_d(m). \eqnd\etadepole $$
Below, we need the
first terms of the expansion of $\Omega_d(m)$ for $m^2\to 0$.
Introducing the cut-off dependent constant
$$\rho_c={1\over(2\pi)^3}\int^\Lambda {\d^3 p\over p^2}
=\Omega _3(0)\,,\eqnd\eRciiidef $$ we rewrite Eqs.~\eSUSYii{} as
\eqna\eSUSYiv
$$\eqalignno{\rho-\varphi^2/N&=\rho_c-{1\over4\pi } m\,, &\eSUSYiv{a}\cr
s-2F\varphi/N &={1\over2\pi}M\left(|M|- m\right). &\eSUSYiv{b}\cr}
$$ Note that, here, the change $U\mapsto -U$ corresponds to
$$F\mapsto -F\,,\quad s\mapsto -s\,,\quad M\mapsto -M\,.$$
\medskip Finally, we calculate the action  density $\cal E$ corresponding
to the action ${\cal S}_N$ (Eq.~\eactSUSYN),  ${\cal E}={\cal
S}_N/{\rm volume}$, ~for vanishing fermion fields:
$$\eqalignno{{\cal E}/N&=-\ud F^2/N+\ud s U'(\rho)+MF\varphi/N+\ud
\lambda\varphi^2/N-\ud Ms -\ud \lambda \rho \cr&\quad
+\half\tr\ln(-\partial^2 +M^2+\lambda) -
\half\tr\ln(\sla{\partial}+M).&\eqnd\eSUSYV }$$
In $d=3$, we have in Eq.~\eSUSYV:
$$    \half\tr\ln(-\partial^2
+M^2+\lambda) - \half \tr\ln(\sla{\partial}+M)  = \ud
\rho_c\lambda-{1\over12\pi}\left(m^3-|M|^3\right) .
\eqnd\eSupertraceLN  $$
Note that the saddle point equations are recovered, in
component form, by taking the derivatives of $\cal E$ with
respect to the various parameters. \par
Using the saddle point Eqs.~\eSUSYsad{a}, \eSUSYiv{} and \eSUSYsadb{a} to eliminate
$F,s,\rho$, one eliminates also the explicit dependence on the
super-potential $U$ and the expression simplifies into
$${\cal E}/N=\ud M^2\varphi^2/N+  {1\over 24\pi}(m-|M|)^2(m+2|M|)
.\eqnd\eSUSYground
$$
In this form we see that the action density $\cal E$ is positive
for all saddle points, and, as a function of $m$, has an absolute
minimum at $m=|M|$, and thus $\lambda =0$,  that is, for a
supersymmetric ground state. \par Eq.~\eSUSYsad{}\  implies
$M\varphi=0$, and thus ${\cal E}=0$. Therefore, if a
supersymmetric solution exists, it will have the lowest possible
ground state energy  and any non-supersymmetric solution will have
a higher energy. Since $M$, $m$, and $\varphi$ are related by the
saddle point equations, it remains to verify whether such a
solution indeed exists.\par In the supersymmetric situation the
saddle point equations reduce to $s=0$, $M\varphi=0$ and
$$\rho-\rho_c=\varphi^2/N-|M|/4\pi\,,\quad M=U'(\rho).$$
In the $O(N)$ symmetric phase $\varphi=0$ and $|M|=4\pi
(\rho_c-\rho)=|U'(\rho)|$. In the broken phase $U'(\rho)=0$ and
$\varphi^2/N=\rho-\rho_c$. We will show in the next section that
these conditions can be realized by a quadratic function $U(\rho)$
and, then, in both phases the ground state is supersymmetric and
$\cal E$ vanishes.
\subsection The $\Phi^4$ super-potential in $d=3$: phase structure

We now consider the special case
$$U(R)=\mu R+\ud u R^2\ \Rightarrow\ U'(R)=\mu+ u R \,.\eqnd\eURsquar$$

The dimensions of the $\theta$ variables and the field $\Phi$ are
$[\theta]=-\ud\,,\quad[\Phi]=\ud\ \Rightarrow \ [u]=0\,.$ Power
counting thus tells us that the model is renormalizable in three
dimensions. Prior to a more refined analysis, one expects coupling
constant and field renormalizations (with logarithmic divergences)
and a mass renormalization with linear divergences. Using the
solution for the two-point function $W^{(2)}$, in Eq.~\eqns{\eSUSYiipt}
in the appendix, one infers that the coefficient $A(p^2)$ has at
most a logarithmic divergence, which corresponds to the field
renormalization, while the coefficient $C(p^2)$ can have a linear
divergence which corresponds to the mass renormalization. \par For
the quartic potential, Eqs. ~\eSUSYsadb{} are now
$$M=\mu+u \rho \,,\quad \lambda=us\,.\eqnd\eSUSYmpot $$
We introduce the critical value of $\mu$:
$$\mu_c=-u \rho_c  \, . \eqnn $$
Taking into account Eqs.~\eSUSYmpot, one finds that Eqs.
~\eSUSYiv{} can now be written as  \eqna\eSUSYivquartic
$$\eqalignno{ M&=\mu-\mu_c + u\varphi^2 /N- {u \over{4\pi}}
\sqrt{M^2+\lambda}\,,&\eSUSYivquartic{a}\cr \lambda &=2u
M\varphi^2/N + {u \over{2\pi}}M\left( |M|-\sqrt{M^2+\lambda}
\right)\,. &\eSUSYivquartic{b}\cr}$$
 Eqs.~\eSUSYivquartic{} relate the fermion
mass $m_\psi=|M|$, the boson mass $m_\varphi = \sqrt{M^2+\lambda}$
and the classical field $\varphi $. The phase structure of the
model is then described by the lowest energy solutions of these
equations in the $\{ \mu-\mu_c , u \}$ plane. The invariance under
the change $U\mapsto -U$ that was mentioned above and seen now in
Eq.~\eURsquar~and the equations that followed. This invariance is
reflected into the phase structure of the model and one can
restrict the discussion to $u>0$. The phase structure is
summarized
 in Fig. \label{\phases} in the first and
 second quadrant of the $ \{ \mu-\mu_c, u/u_c \} $.
\par
We find, indeed, that supersymmetry is left unbroken ($\lambda =
0$) and the ground state energy ${\cal E}=0$ in each quadrant in
the $\{ \mu-\mu_c , u \}$ plane. This is consistent with
Eqs.~\eSUSYivquartic{} having a common solution with $\lambda =0$
(thus $m_\psi= m_\varphi = |M| $) and $M\varphi=0$. They then
reduce to
$$ M =\mu-\mu_c + u{\varphi^2 \over N} - {u \over{4\pi}}|M|\,,
\quad M\varphi  =0\, . \eqnd\eSUSYiii $$
\medskip
{\it The broken $O(N)$ symmetry phase:} The  $M=0$ solution
implies a spontaneously broken $O(N)$ symmetry, scalar and fermion
$O(N)$ quanta are massless and
$$\varphi^2 =-N(\mu-\mu_c)/u\,, \eqnd\ephi$$
which implies that this solution exists only for $ \mu<\mu_c$. The
solution exists in the fourth (and second) quadrant of the $\{
\mu-\mu_c , u \}$ plane.
\medskip
{\it The $O(N)$ symmetric phase:} ~Here, $\varphi^2=0$ and the
equation
$$M=\mu-\mu_c - (u/u_c)|M| \eqnd\eSymmetricPhase$$
yields the common mass $M$ for the fermions and bosons. In
Eq.~\eSymmetricPhase\ we have introduced the special value $u_c$
of the coupling $u$:
$$u_c= 4\pi  \ . \eqnd\egCritical$$
Eq.~\eSymmetricPhase\ splits into two equations, depending on the sign of
$M$. The first solution
$$M= M_+=(\mu-\mu_c)/(1+u/u_c) >  0\,,$$
exists only for $\mu>\mu_c$ (first quadrant), as one would
normally expect.  \par
The second solution
$$M=
M_-=(\mu-\mu_c)/(1-u/u_c) < 0 \,,$$
is very peculiar. There are
two different situations depending on the position of $u$ with
respect to $u_c$: \par
 (i) $u>u_c=4\pi$ and then $\mu>\mu_c$:  the
solution is degenerate with another $O(N)$ symmetric solution
$M_+$.\par
 (ii) $u<u_c$ and then $\mu<\mu_c$: the solution is
degenerate with a solution of broken $O(N)$ symmetry. \par
\midinsert \epsfxsize=13.5cm \epsfysize=11cm \pspoints=2.pt \vskip
-5.5cm \hskip2cm \epsfbox{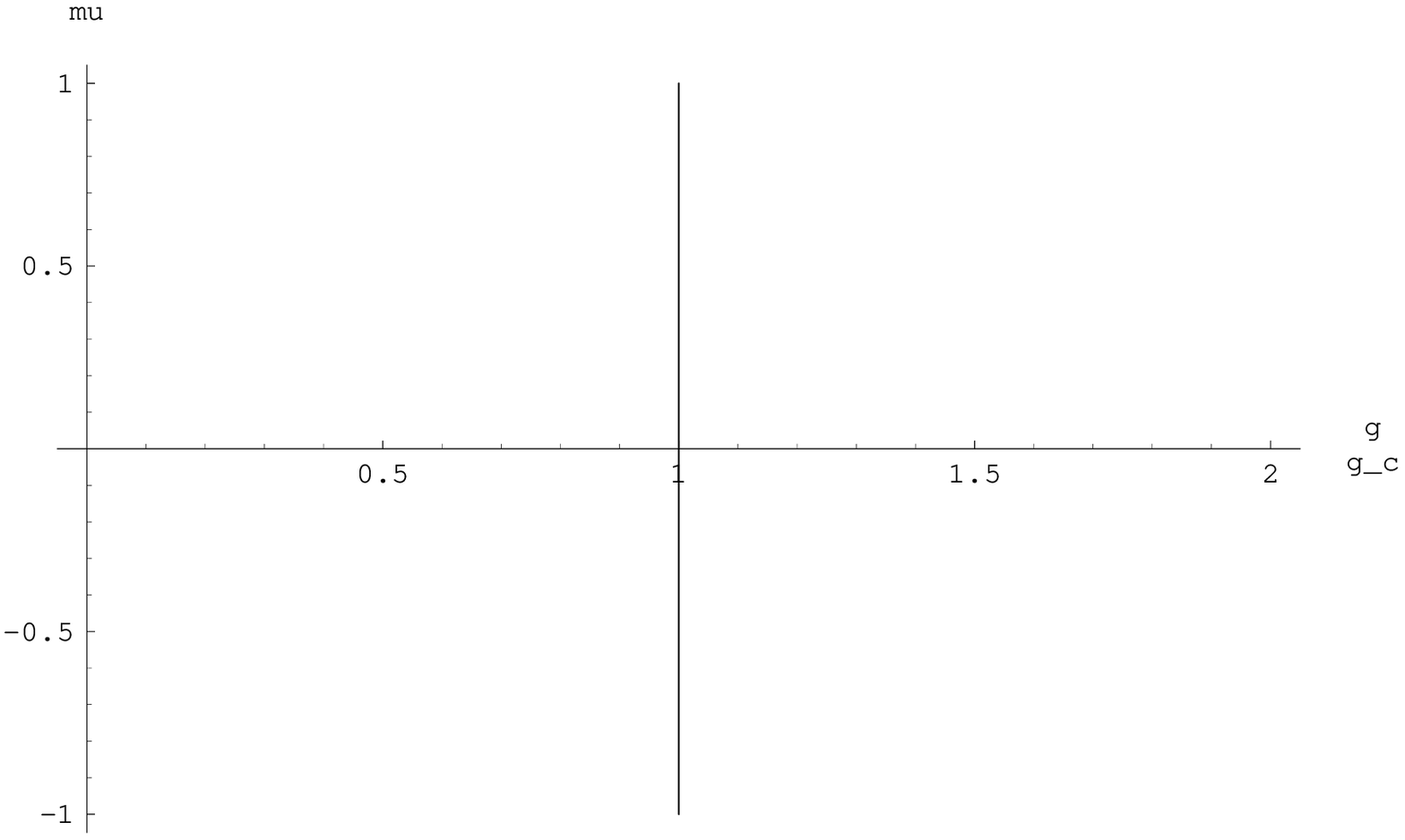}
\vskip -6.cm
\hskip 4.7cm $\varphi^2 = 0$ \vskip .4cm \hskip 2.05cm
$~~~~\mu-\mu_c$ \vskip .1cm \hskip 3.5cm {\bf I} \hskip 3cm {\bf
II} \vskip .1cm \hskip 7cm $M_-$ \vskip .2cm \hskip 3.3cm $M_+$
\hskip 3cm $M_+$
\vskip 0.15cm
 \hskip 8.3cm $u/u_c$ \vskip .3cm \hskip 3.2cm $M_-$
\vskip .3cm \hskip 3.5cm {\bf IV} \hskip 3cm {\bf III} \vskip 1cm
\hskip 5cm $\varphi^2 \neq 0$ \vskip 1cm \figure{1mm}{Summary of
the phases of the model in the $ \{ \mu-\mu_c , u \} $ plane.
Here  $\ma=\mpsi=|M_\pm|=(\mu-\mu_c)/(u/u_c\pm 1)$. The lines
$u=u_c$ and $\mu-\mu_c=0$ are lines of first and second order
phase transitions.} \figlbl\phases
\endinsert
The following different phases appear in the phase structure
 that is summarized in Fig. \label{\phases}:
\smallskip
{\bf Region I : ~~$\mu-\mu_c \geq 0 ~~,~~ {u/ u_c}\leq 1$}:

\noindent Here, there is only one $O(N)$-symmetric, supersymmetric
ground state with $\mpsi=\ma=M_+=(\mu-\mu_c)/(u/u_c+1)$ and
$\varphi^2=0$.
\smallskip
{\bf Region II : ~~$\mu-\mu_c \geq 0 ~~,~~ {u/ u_c}\geq 1$}:

\noindent There are two degenerate $O(N)$-symmetric
($\varphi^2=0$), supersymmetric ground states with masses
$\mpsi=\ma=M_+=(\mu-\mu_c)/(u/u_c+1)$ and
$\mpsi=\ma=-M_-=(\mu-\mu_c)/(u/u_c-1)$.

\smallskip

{\bf Region III : ~~$\mu-\mu_c \leq 0 ~~,~~ {u/ u_c}\geq 1$}:

\noindent There is one supersymmetric ground state, it is an
ordered state with broken $O(N)$ symmetry ($\varphi^2 \neq 0$,
$\mpsi=\ma=0$).
\smallskip

{\bf Region IV : ~~$\mu-\mu_c \leq 0 ~~,~~ {u/ u_c}\leq 1$}:

\noindent There are two degenerate ground states: An $O(N)$-symmetric ,
 supersymmetric ground states with masses
$\mpsi=\ma=M_-=(\mu-\mu_c)/(u/u_c-1)$ and $\varphi^2=0$. The
second ground state is a supersymmetric, broken $O(N)$ symmetry
state with $\mpsi=\ma=0$ and $\varphi^2 \neq 0$.
\medskip

{\it The action density:} In order to exhibit the phase structure
in terms of the variation of the  action density ${\cal E}$, we
plot in Fig. \label{\figTIVaa }~~the expression \eSUSYground, but
use only the fermion gap equation in Eq.~\eSUSYivquartic{a}, in
such way that $\cal E$ remains a function of $\varphi$ and
$\lambda $, or equivalently $\varphi$ and $m=\sqrt{M^2+\lambda }$:
$$  {1\over N} {\cal E}(m,\varphi)
 = \half M^2(m,\varphi ) {\varphi^2 \over N}
 +{1\over 24\pi} \left[m -\left|M(m,\varphi )\right|\right]^2
  \left(\ma + 2\left|M(m,\varphi)\right| \right) .
 \eqnd\HartreeVTzeroE $$

\midinsert \epsfxsize=18cm \epsfysize=14cm \pspoints=2.pt \vskip
-7.cm \hskip1.5cm \epsfbox{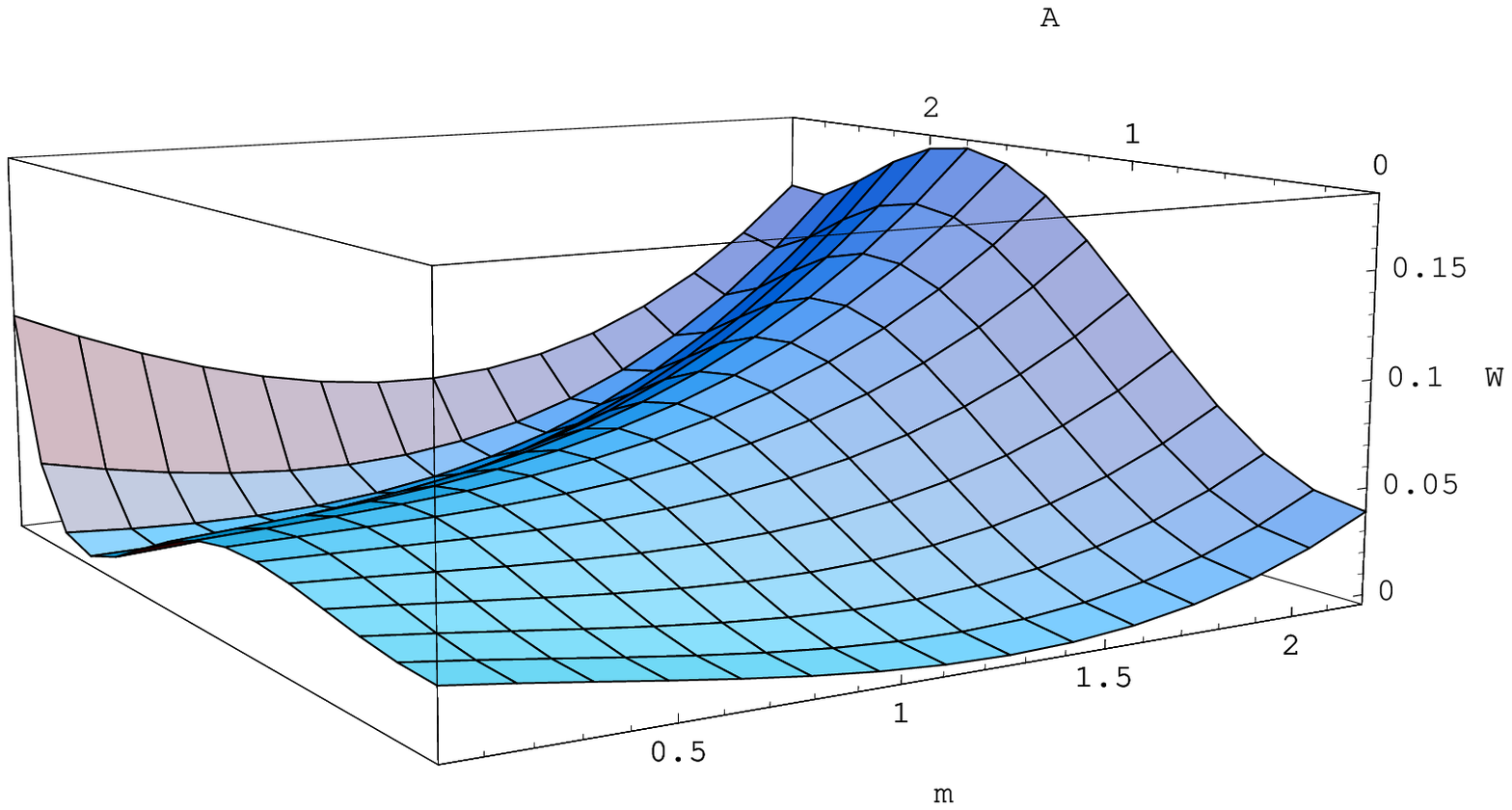} \figure{1mm}{{\bf Region
IV :} The  energy density $W(m,\varphi )\equiv {1\over N} {\cal
E}(m ,\varphi )$ as given in Eq.~\HartreeVTzeroE\ as a function of
the boson mass ($m$) and $A$, where $A^2=\varphi^2/u_c$. Here
$\mu-\mu_c=-1 ~,~ u/u_c=0.2 $. As seen here  there are two
distinct degenerate phases. One is an ordered phase ($\varphi \neq
0$) with a massless boson and fermion, the other is a symmetric
phase ($\varphi = 0$) with a massive ($m=|M_-|$) boson and
fermion.
 } \figlbl\figTIVaa
\endinsert

Fig.~\label{\ZeroTone} displays the action density  $\cal E$  restricted
to $\varphi=0$:
$$  {1\over N} {\cal E}(m,\varphi = 0)
=  {1\over 24\pi} \left[m - \left| \mu -\mu_c - (u/u_c) m \right|
\right]^2   \left( m +2\left| \mu -\mu_c -(u/u_c)m \right|
\right).   \eqnd\HartreeVTzeroF  $$

\midinsert \epsfxsize=13.5cm \epsfysize=10cm \pspoints=2.pt \vskip
-5cm \hskip 2cm \epsfbox{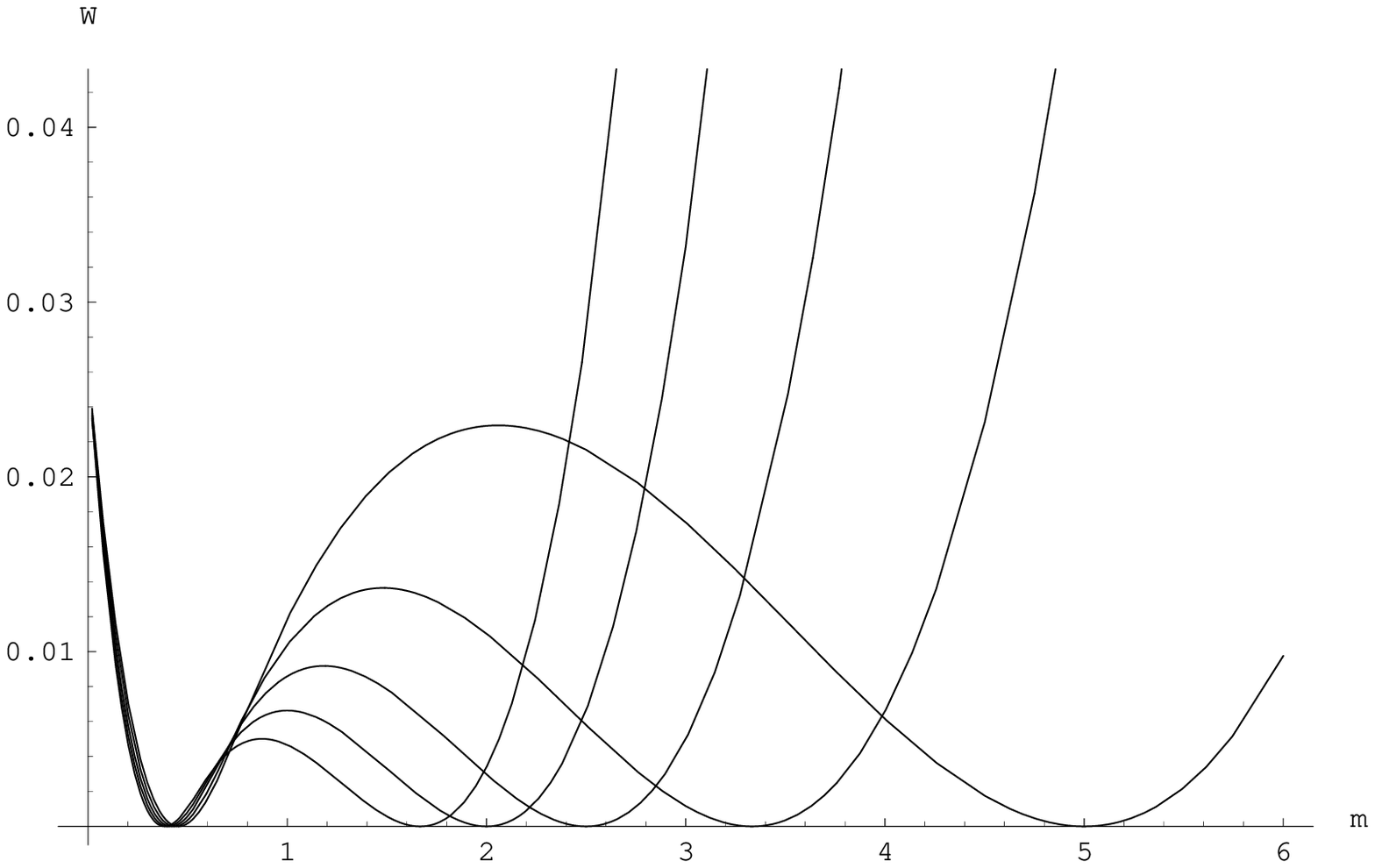} \figure{1mm}{The action
density $W(m)\equiv {1\over N} {\cal E}(m,\varphi=0) $ from
Eq.~\HartreeVTzeroF\  in region {\bf II} of Fig.~\phases. Here
$\mu-\mu_c=1 $ (sets the mass scale) and $u/u_c$= varies between
1.6 and 1.2. There are two massive, degenerate $O(N) $ symmetric
SUSY vacua with $\varphi=0 $ and  possibly  $|M_-/M_+|>> 1 $ (as
$u \to u_c $ ) with $m_\psi=m_\varphi=M_+=(\mu-\mu_c)/(u/u_c+1)$
and $m_\psi=m_\varphi=-M_-=(\mu-\mu_c)/(u/u_c-1)$. }
\figlbl\ZeroTone
\endinsert

Several peculiar phase transitions can be easily traced now in
Eq.~\HartreeVTzeroE{}. First, one notes the phase transitions that
occur  when $\mu-\mu_c$ changes sign. When  $ 0 < u <u_c  $ and
$\mu > \mu_c$, the system has a non-degenerate $O(N)$-symmetric
ground state with bosons and fermions of mass $M=M_+$. As $\mu -
\mu_c$ changes sign ($0 < u <u_c$ fixed), two degenerate  ground
states appear, as seen in Fig.~\figTIVaa . Either $M=0$ and
$\varphi^2=-N(\mu-\mu_c)/u$ or the system stays in an
$O(N)$-symmetric ground state with a mass $|M_-|$ for the bosons
and fermions. Similarly, when one goes from $\mu < \mu_c$ to $\mu
> \mu_c$ at $ u>u_c= 4\pi  $, the $O(N)$ symmetry is restored but
there are two degenerate  ground states to choose from,
$M=M_\pm$.


\midinsert \vskip -3.5cm \epsfxsize=14.5cm \epsfysize=8cm
\pspoints=2.pt \hskip 2cm \epsfbox{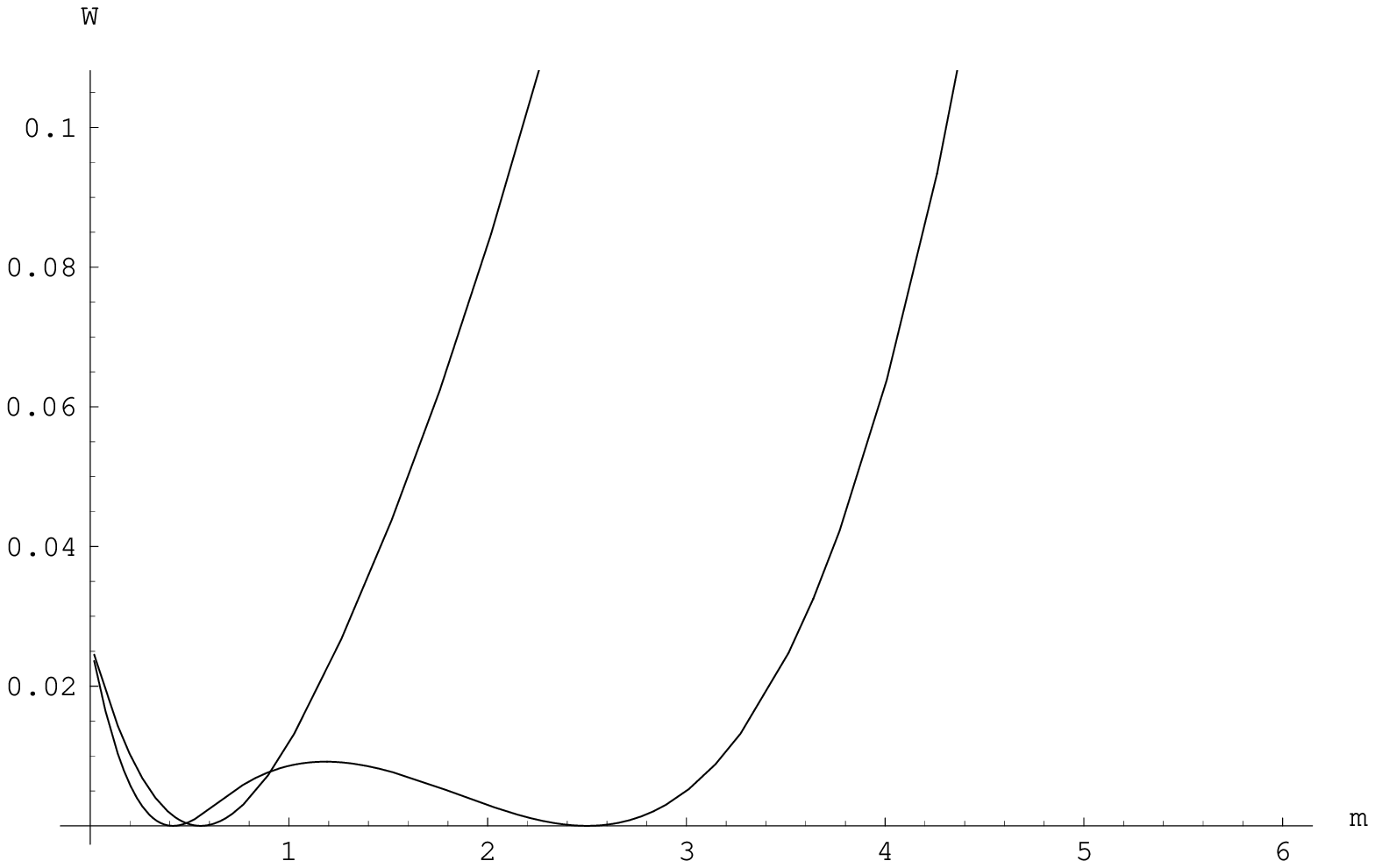} \figure{1mm}{As in
Fig.\ZeroTone, ~the action density $W(m)\equiv {1\over N} {\cal
E}(m,\varphi=0)$ with $\mu-\mu_c=1 $ and $u/u_c = 1.2$ is changed
to $u/u_c = 0.8$ (from region {\bf II} to region {\bf I} of
Fig.~\phases). There are two degenerate $O(N)$ symmetric SUSY
vacua   at $u/u_c= 1.2$ with masses  $m_\psi=m_\varphi=|M_\pm|$
where $|M_\pm|=\mu/(u/u_c \pm 1)$ while at $u/u_c= 0.8$ there is a
non-degenerate vacuum at $m_\psi=m_\varphi=M_+$.} \figlbl\ZeroTtwo

\endinsert

 As seen  in Figs.~\label{\ZeroTone} and
\label{\ZeroTtwo}, an unusual transition takes place when one
varies the coupling constant $u$. For positive $\mu-\mu_c$ we find
two degenerate ground states if $ u
> u_c $. As $u$ is lowered
(at fixed $\mu-\mu_c$) the ground state with mass $\ma=\mpsi=-M_-$
disappears ($|M_-/ M_+| \to \infty$) and only the $O(N)$-symmetric
phase remains with $M=M_+$.
Namely, suppose we consider at $\{ u>u_c$, $\mu-\mu_c>0\}$ a
physical system in a state denoted by $A$ and defined by
$\{\varphi^2=0, M=M_-\}$. Such a system will go into a state $B$
defined by $\{\varphi^2=0, M=M_+\}$ when $u$ {\bf decreases} and
passes the value $u=u_c= 4\pi $. Now, if we consider the reversed
process; a physical system at $\{ u < u_c , \mu-\mu_c> 0 \}$ that
is initially in the ground state $B$ and $u$ {\bf increases }and
 passes  $u=u_c$. There is now no reason  for the
system to go through  the reversed transition from $B$ to $A$
since the $O(N)$-symmetric states with $M=M_-$ and $M=M_+$ are
degenerate and since supersymmetry is preserved the energy of
state $A$ will not decrease below zero.  These peculiar phase
transitions with ($|M_-/ M_+| \to \infty$) and with ``infinite
hysteresis" in the $A \to B$ transitions  are due to the fact that
supersymmetry is left unbroken and the degeneracy of the two
possible ground states in the leading order in $1/N$. If
supersymmetry would have been broken by some small parameter, the
lifted degeneracy would, most probably, have translated into a
slow first order transitions between the otherwise degenerate
ground states.

The transition from the degenerate vacua at $u/u_c =1.2$ to a
non-degenerate  ground state at $u/u_c= 0.8$ is shown in
Fig.~\label{\ZeroTtwo}  (from region {\bf II} to region {\bf I}).
This type of transitions between the different phases of
~Fig.~\phases~ will be studied as a function of the temperature in
the next section.
\medskip
{\it Special situation.} In general, when $\mu=\mu_c$ the mass $M$
vanishes. However, there is a special case when
$$u=u_c=4\pi  \,. $$
Then the value of $M$ is left undetermined. An accumulation point
of coexisting degenerate ground states exist in the phase
structure shown in Fig.~\phases. The case $\mu=\mu_c$ represents a
scale invariant theory where, however, the $O(N)$ fermionic and
bosonic quanta can have a non-vanishing mass $\ma=\mpsi=|M|$.
Since $u$ has not gone any perturbative renormalization there is
no explicit breaking of scale invariance at this point. Thus, the
only scale invariance breaking comes from the solution of the gap
equation for $M$ which leaves, however, its numerical value
undetermined. We see a dimensional transmutation from the
dimensionless coupling $u$ that is fixed at a value of $u=u_c$
into an undetermined scale $M$. If $ M \neq 0$ the spontaneous
breaking of scale invariance will require the appearance of a
Goldstone boson at the point $ u= u_c$. The massless Goldstone
boson is associated here with the spontaneous breaking of scale
invariance~\BHM. Moreover, since the ground state is
supersymmetric, we expect the appearance of a massless $O(N)$
singlet Goldstone boson (a dilaton) and its massless fermionic
partner (a ``dilatino''). In order to verify these properties, we
now calculate the $\left<LL\right>$ propagator that will enable us
to see these poles in the appropriate four-point functions.

\subsection $\left<LL\right>$ propagator, bound states and massless fermion and boson $O(N)$ singlet bound states

{\it The $\left<LL\right>$ propagator.} We now calculate the
$\left<LL\right>$ propagator in the symmetric phase. Then, the
propagators of the superfields $\phi$ and $L$ are decoupled. In
the example  of a quartic potential,  the $\rho$ field can be
eliminated by gaussian integration. The relevant part of the
$L$-action then reads
$$-{N\over 2u}\int\d^3 x\,\d^2\theta(L-\mu)^2+\ud (N-1) {\rm Str}\,
\ln\left(-\bar \D  \D +2L\right) . $$
The calculation of the
$\left<LL\right>$ propagator  involves the super-propagator
\esupprop. For the inverse propagator $\Delta_L^{-1}$ one finds
$$\Delta_L^{-1}(p)=-{N\over u}\delta^2(\theta'-\theta)
-2N\int{\d^3 k\over(2\pi)^3}
\Delta(k,\theta,\theta')\Delta(p-k,\theta,\theta')   $$ with,
here, (see Eq.~\eSUSYiipt)
$$\Delta(k,\theta,\theta')={1\over k^2+M^2}\left[1+\ud
M\delta^2(\theta'-\theta) \right]\e^{i \bar\theta
\slam{k}\theta'}\,. $$
Then,
$$\Delta(k,\theta,\theta')\Delta(p-k,\theta,\theta')
={[1+M \delta^2(\theta'-\theta)]\e^{i \bar\theta \slam{p}\theta'}
\over (k^2+M^2)[(p+k)^2+M^2]}\,. $$
Notice the cancellation of the
factor $\e^{i \bar\theta \slam{k}\theta'}$ which renders the
integral more convergent that one could naively expect. The
integral over $k$ then yields the three-dimensional bubble diagram
$B(p)$:
$$\eqalign{B(p)={1\over(2\pi)^3}\int{\d^3 k\over  (k^2+M^2)[(p+k)^2+M^2]}
={1\over4\pi p}{\rm Arctan}(p/2|M|)\,. \cr}
$$
At leading order for $p$ small, we need only $B(0)=1/8\pi |M|$.
Then,
$$\Delta_L^{-1}=
-{N\over4\pi |M|}\left[1+\{M +|M|(u_c/u) \}
\delta^2(\theta'-\theta)\right]\e^{i \bar\theta \slam{p}\theta'}.
$$
Considering that in momentum space we have
$$[-\bar {\rm D} {\rm D}+2\mu]\delta^2(\theta'-\theta)=
 2\left[-2+ \delta^2(\theta'-\theta)\mu  \right]
 \e^{-i\bar\theta\slam{k}\theta'} ,\eqnd\eSUSYker $$
 we conclude that for small $M>0$ the $LL$
propagator corresponds to a super-particle of mass $2M (1+u_c/u )
$. For $M<0$ the mass is $2|M(1-u_c/u)|$. For $|u-u_c|$ small, it
is a bound state and at the special point $u=u_c$ the mass
vanishes.\par More generally, we find
$$\Delta_L(p)={2\over NB(p)}
{1\over p^2+m^2(p)}\left[1-\ud m(p)\delta^2(\theta'-\theta)\right]
\e^{i \bar\theta \slam{p}\theta'}  \eqnd\eLLprop$$
with
$$m(p)=2M+{1\over  u B(p)}\,.$$
We note that the only renormalization required at leading order is
a mass renormalization, a situation similar to the $\varphi^6$
scalar field theory~\BBM. As a consequence field dimensions are
not modified.\par

Clearly, the propagation of  the fields $M(x)$  and $\ell(x)$, as
indicated in Eq.~\eLLprop , when combined with the $
L(\theta)\Phi^2$ interaction in Eq.~\eactLagm, namely,
$$  \int\d^2\theta\,L\Phi^2 =M(-\half\bar\psi\psi+F\varphi )
-\varphi\bar\ell \psi+\half\lambda \varphi^2, $$ describes the
bound states in the $\varphi\varphi$, $\psi\psi$ and $\psi\varphi$
scattering amplitudes. For example~\BHM, in the supersymmetric
ground state case and with $\mu-\mu_c=0 $, the $\psi\varphi$
scattering amplitude $T_{\psi\varphi,\psi\varphi}(p^2)$, in the
limit $p^2 \to 0$ satisfies
$$T_{\psi\varphi,\psi\varphi}(p^2) \sim {2u\over
N}\left[1+ {u\over 4\pi}{M\over |M|} +{u\over 2\pi}{i\sla{p} \over
|M|} \right]^{-1}  \to -{4\pi i\over N}{|M|\over \sla{p}} \eqnn $$
for  $M<0$ and $u \to u_c$

One notes here that the fermion massless bound state pole appears
when a non-zero solution  ($M$) to the gap equation exists $
(m_\varphi=m_\psi=|M|) $ in the absence of any dimensional
parameters ($\mu-\mu_c=0$).  This happens  when the force between
the massive $\psi$ and $ \varphi$ quanta is determined by $u \to
u_c$. The massless $O(N)$ singlet fermion bound state excitation
is associated with the spontaneous breaking of scale invariance.
Similarly, the bosonic partner of this massless bound state
excitation can be then seen, at the same value of the parameters,
in $\varphi\varphi$ and $\psi\psi$ scattering amplitudes as
 Eq.~\eLLprop\ shows.

At $\mu=\mu_c$ in the generic situation $M=0$, or for
$|p|\to\infty$ we find
$$B(p)={1\over 8p} $$
and, thus,
$$\Delta^{-1}_L(p)=-{N\over u}\delta^2(\theta'-\theta)
-{N\over4 p}\e^{i \bar\theta \slam{p}\theta'}. $$
As a consequence
the canonical dimension of the field $L$ is one, as in
perturbation theory, and the interaction $L\Phi^2$ in
Eq.~\eactLagm\ is renormalizable.

\subsection  Dimensions $2\leq d \leq 3$

In $d$ space-time dimensions $2\leq d \leq 3$, the space of
$\gamma $ matrices implies  $\tr{\bf 1}=2$ and one notices, after
some calculations following the same pattern as discussed above,
that Eqs.~\eSUSYii{} are replaced by
\eqna\esadNSUSYnls
$$\eqalignno{\rho-\varphi^2/N&=\Omega _d(m), &\esadNSUSYnls{a}\cr
s-2F\varphi/N&=2M\left[\Omega _d(m)-\Omega _d(|M|)\right]. \cr}$$
Taking into account the other saddle point equations and
$\rho_c=\Omega _d(0)$, one obtains a generalization of
Eq.~\eSUSYground:
$$2 {\cal E}/N =  M^2\varphi^2/N+(M^2-m^2)\left[ \Omega _d(m)-\Omega _d(0)\right] +\int_0^\lambda \d\lambda ' \,
 \Omega _d\left(\sqrt{M^2+\lambda '}\right)  \eqnd \eDminEi $$
and, thus,
$$ {\cal E}/N =\half M^2\varphi^2/N +K(d)\left[ \ud (m^2-M^2)m^{d-2}
-\left(m^d-|M|^{d }\right)/d\right],$$
 where
$$K(d)= -{ \Gamma(1-d/2)\over (4\pi)^{d/2}}\,.$$
Moreover, from Eq.~\eDminEi,
$${\partial {\cal E}\over\partial m }=- {N\over 2m}  (m^2-M^2)
 \Omega _d'(m) \eqnd \eDminE $$
has the sign of $m-|M|$ because $\Omega _d'(m)$ is negative.
The function, therefore, has a minimum  at $m=|M|$ for all values of
$d$, and supersymmetry is maintained in the ground state for all
$2\leq d \leq3$.\par
\medskip
{\it Two dimensions.} It may be interesting to consider the same
model in two dimensions. The model now is super-renormalizable.
Even at $\mu=0$ it is not chiral invariant since a chiral
transformation that changes $\bar\psi\psi$ in $-\bar\psi\psi$, is equivalent to the change $U\mapsto -U$.\par The
expression \eDminE\  is cut-off independent and has a limit for
$d=2$:
 $${\cal E}/N=\ud M^2\varphi^2/N +{1\over8\pi}
\left[m^2-M^2-2M^2\ln(m/M)\right]  ,\eqnd\eSUSYNEnii $$ an
expression which again has an absolute minimum at $m=|M|$. Then,
a minimization with respect to $M$ and $\varphi$ yields $M\varphi=0$.
At the minimum $\cal E$ vanishes. \par
Taking into account
$\lambda =0$ and Eq.~\esadNSUSYnls{a} in the $d=2$ limit, we obtain
the gap equation
$$M=\mu+u\varphi^2/N+{u\over 2\pi}\left[\ln(\Lambda /|M|]+O(1)\right). $$
The solution $M^2=\lambda=0$, therefore,  is not acceptable. The
$O(N)$ symmetry is never broken. When the mass $M$ is small in the
cut-off scale, the l.h.s.~is negligible and
$$m=M\propto  \Lambda  \e^{2\pi\mu /u}. $$
This solution exists only when $-\mu/u$ is positive and large.

\subsection A supersymmetric non-linear $\sigma $-model at large $N$

We now consider the supersymmetric non-linear $\sigma $-model in
$d$ dimensions, $2\le d\le 3$. The action  \sslbl\ssSUSYnls
$${\cal S}(\Phi)={1\over 2 \kappa }\int\d^d x\,\d^2\theta\, \bar \D \Phi\cdot
\D \Phi   \eqnd\eNLSM $$ involves a scalar superfield $\Phi$,
an $N$-component vector that satisfies
$$\Phi\cdot\Phi = N \,.$$
This relation is implemented  by introducing  a superfield
$$L (x,\theta) = M(x) + \bar\theta\ell(x) +\half \bar\theta\theta
\lambda(x) ,$$ where $M(x)$, $\lambda(x)$ and $\ell(x)$ are the
Lagrange multiplier fields, and adding to the action
$${\cal S}_L ={1\over  \kappa }\int\d^d x\,\d^2\theta \, L (x,\theta)
\left[\Phi^2(x,\theta)-N \right] . \eqnd\eNLSML $$
The partition
function is given by (${\cal S}(\Phi,L)={\cal S}+{\cal S}_L$)
$${\cal Z} = \int [d\Phi][\d L] \e^{ -  {\cal S  }(\Phi,L) }.$$
In terms of component fields, the total action reads
$$\eqalignno{ {\cal S}={1\over  \kappa }\int \d^dx \{ &\half { \varphi}(-\del^2 +
M^2){ \varphi} -\half{ {\bar\psi}}(\delslash+M){ \psi} \cr &+\ud
\lambda ({ \varphi}^2- N  ) -\bar\ell({ {\bar\psi}}\cdot \varphi)
\}. &\eqnd\NLSMcomp }$$ As in the case of the $\Phi^4$ theory, we
integrate out $N-1$ superfield components leaving out
$\phi=\Phi_1$:
$${\cal Z} = \int [d\phi][\d L] \e^{ - {\cal S}_N(\phi,L) },$$
where
$$   {\cal S}_N(\phi,L)= {1\over\kappa } \int\d^d x\,\d^2\theta  \left[ \half\bar \D \phi\cdot
\D \phi+  L \left(\phi^2- N \right) \right]    + {{N-1}\over 2} {\rm
Str}\, \ln( -{\bar \D }\D  +2 L )  .   \eqnd\eNLSMeff   $$
By varying
the effective action with respect to $\phi$, one finds the saddle
point equation
$${\bar \D }\D \phi-2L \phi =0 \,,\eqnd\eSadNLSMa $$
which implies, for constant $\varphi$ and $\psi=0$,
$$ F-M\varphi=0 \qquad {\rm and} \qquad MF+\lambda\varphi = 0 \,.
\eqnd\eSadNLSMaa$$
Varying the large $N$ action  with
respect to $L $ and using the expression  \esupprop~for the
$\phi$ propagator, one finds ($N\gg 1$)
$${1\over \kappa}-{\phi^2 \over N}
=  \int{\d^d k\over (2\pi)^d}\left[{1+ M\bar\theta\theta \over
k^2+M^2+\lambda}-{M\bar\theta\theta\over k^2+M^2}\right].
\eqnd\eSadNLSMb$$
We now introduce the boson mass parameter
$m=\sqrt{M^2+\lambda }$. In terms of its components
Eq.~\eSadNLSMb~yields (for  $\psi=0$)
 \eqna\eSadNLSMbb
$$\eqalignno{1-{\varphi^2\over N}&=
\kappa \,\Omega _d (m ), &\eSadNLSMbb{a}\cr M{\varphi^2\over
N\kappa } &=M \left[\Omega _d(|M|)- \Omega _d\left( m\right)
\right]. &\eSadNLSMbb{b}\cr} $$
\medskip
{\it Dimension $d=3$.} Introducing the critical (cut-off
dependent) value $\kappa _c$:
 $${1\over\kappa_c}= \Omega _3(0)\,,$$
we can write Eqs.~\eSadNLSMbb{} now as \eqna\eSadNLSMbbb
$$\eqalignno{{1\over \kappa }-{1\over \kappa_c }-{\varphi^2\over N\kappa } &=-
{m\over4\pi }\,, &\eSadNLSMbbb{a}\cr M{\varphi^2\over N\kappa }
&={1\over4\pi}M\left(m-|M|\right). &\eSadNLSMbbb{b}\cr} $$ The
calculation of the ground state energy density of the non-linear
$\sigma $-model follows the similar steps as in the $(\Phi^2)^2$
model. One finds the action density
$$  {\cal E} /N={ 1\over 2N\kappa } m^2\varphi^2  - \ud(m^2-M^2)\left( {1\over\kappa } -{1\over\kappa_c}\right)
-{1\over12\pi}\left(m^3-|M|^3\right)
   \eqnd\eNLSMenergy $$
and, taking into account the saddle point Eq.~\eSadNLSMbb{a},
 $$ {\cal E} /N={ 1\over 2N\kappa } M^2\varphi^2 +  {1\over 24\pi}(m-|M|)^2(m+2|M|)
,\eqnd\eSUSYnlsground
$$
an expression identical to \eSUSYground, up to the normalization
of $\varphi$. Again, if a supersymmetric solution exists it has the
lowest energy. We thus look for supersymmetric solutions.\par
In the $O(N)$ symmetric phase ($\varphi = 0$), Eq.~\eSadNLSMbbb{b} is
satisfied while Eq.~\eSadNLSMbbb{a} yields
$$m=|M|=4\pi\left({1\over \kappa _c}-{1\over \kappa }\right).$$
This phase exists for   $\kappa \ge \kappa _c$.\par In the broken
phase $m=M=0$, and Eq.~\eSadNLSMbbb{b} is again satisfied.
Eq.~\eSadNLSMbbb{a} then  yields
$$\varphi^2/N=1-\kappa /\kappa _c \,, \eqnn $$
which is the solution for $  \kappa \le \kappa _c$.\par
Since we
have found solutions for all values of $\kappa $, we conclude that
the ground state   is always supersymmetric. At $\kappa _c$, a
phase transition occurs between a  massless phase with broken $O(N)$
symmetry  ($\varphi^2 \neq 0$)  and a symmetric phase
with massive fermions and bosons of equal mass. The phase
structure of the supersymmetric non-linear $\sigma $-model is
analogous to the structure of the usual non-linear $\sigma $-model
and less surprising than in the
$(\Phi^2)^2$ field theory.

\medskip
{\it Dimension $d=2$.}   The phase structure of the supersymmetric
non-linear \hbox{$\sigma$-model} in two dimensions is rather simple, and
again analogous to the structure of the usual non-linear $\sigma
$-model.  Eq.~\eSadNLSMbbb{a} immediately implies that $M^2+\lambda
=0$ is not a solution and thus the $O(N)$ symmetry remains
unbroken $(\varphi^2=0$) for all values of the coupling constant
$\kappa $. Then, with a suitable normalization of the cut-off
$\Lambda $,
$$m=\Lambda \e^{-2\pi/\kappa }. \eqnd\enlsSUSYNii $$
Correspondingly, the action density becomes
$$  {{\cal E}  \over N}=   -  {\lambda\over2\kappa}
 -{1\over8\pi}\left[(M^2+\lambda)\ln(M^2+\lambda)
-\lambda\ln  \Lambda ^2 -M^2\ln M^2 -\lambda\right]
\eqnd\eNLSMenergyiicr $$ and, taking into account Eq.~\enlsSUSYNii,
$$ {{\cal E}  \over N}={1\over 8\pi}\left[m^2-M^2+2M^2\ln(M/m)\right].$$
We recognize expression \eSUSYNEnii, and conclude in the same way
 that a  supersymmetric solution has the lowest energy. Supersymmetry remains unbroken, and the common boson and fermion mass  is
$$M=m=\Lambda \e^{-2\pi/\kappa }. \eqnn $$
The condition $M\ll\Lambda $ implies that the non-trivial physics
is concentrated near $\kappa =0$.


\section  $O(N)$ supersymmetric model  at finite temperature

We now study the phase structure of a supersymmetric model in $d=3$ space--time dimensions at finite temperature.
%
\subsection {The free energy at finite temperature}

We start again from the action \eSUSYact, which is transformed
into
$${\cal S}(\Phi,\rho,L)=\int\d^3 x\,\d^2\theta\left[\ud\bar \D \Phi\cdot
\D \Phi+NU(\Phi^2/N)+L(\theta )(\Phi^2(\theta )-NR(\theta )\right],  $$
$\Phi(\theta,x) , L(\theta,x)$ and $R(\theta,x)$ being $N$-component
scalar superfields, parametrized as in the $T=0$ discussion.

Integrating over $N-1$ superfield components  of $\Phi$ and
keeping $\Phi_1\equiv \phi$ (the scalar component of the
superfield $\phi$ is identified as $\varphi_1 \equiv \varphi$) one
finds an expression for the  partition function ${\cal Z}$ that
differs from Eq.~\eZSUSYfiv\ only by the field boundary conditions
$${\cal Z}=\int[\d\phi][\d R][\d L]\e^{-{\cal
S}_N(\phi,R,L)},\eqnd\Nintegration $$
where ${\cal S}_N$ is the  large $N$ action \eactSUSYN:
$$\eqalignno{{\cal S}_N=\int\d^3 x\,\d^2\theta &\left[\half\bar \D \phi
\D \phi+NU(R)+ L \left(\phi^2-NR\right)\right] \cr &\quad+
\half(N-1){\rm Str}\, \ln\left[-\bar \D
\D +2L\right].& \eqnd\eactSUSYNb \cr}$$
Among the three saddle point equations at $T=0$, only one is changed
at finite temperature:
 \eqna\saddle
$$\eqalignno{ 2L\phi-\bar \D  \D \phi&=0\,,&\saddle{a} \cr
  L-U'(R)&=0\,,&\saddle{b} \cr
 R-\phi^2/N&= {1\over N}\tr\Delta(k,\theta,\theta). &\saddle{c}\cr
} $$ Eq.~\saddle{c}  involves $\Delta(k,\theta,\theta)$
(Eq.~\eSUSYpropcoin)
$$
\Delta(k,\theta,\theta)= {1\over k^2+M_T^2+\lambda}+
\bar\theta\theta M_T \left( {1\over k^2+M_T^2+\lambda} - {1\over
k^2+M_T^2} \right) ,\eqnd\SUSYpropagator$$
where we call $M_T$ the expectation value of $M(x)$ at finite
temperature $T$. \par
When written in components, Eq.~\saddle{a} implies
\eqna\saddleComponA
$$\eqalignno{F-M_T\varphi&=0 \  , &\saddleComponA {a} \cr
 \lambda\varphi+M_TF&=0 \ .&\saddleComponA{b} \cr} $$
and, thus, the Goldstone condition $\varphi(\lambda+M_T^2)=0 $
follows.\par Eq.~\saddle{b} implies
$$M_T=U'(\rho)\ ,\quad  \lambda=sU''(\rho).\eqnd\saddleComponB $$
When calculating the trace in Eq.~\saddle{c}, we have
to take into account
 that bosons at finite temperature satisfy
periodic and fermions anti-periodic boundary conditions. For
bosons we introduce the function
$$\eqalignno{G_2(m_T, T)&  ={T\over (2\pi)^{d-1}  }\sum_{n\in{\cal Z}}\int^\Lambda {\d^{d-1} k\over (2\pi n T )^2+k^2+m_T^2} \cr
 &=\int^\Lambda{\d^{d-1} k\over (2\pi)^{d-1}}{1\over \omega (k)}
 \left({1\over2} +{1\over \e^{\beta  \omega_\varphi (k)}-1}\right) .&\eqnd\eNfivTtp
\cr}$$ with $\omega_\varphi (k)=\sqrt{M_T^2+\lambda +k^2}. $ It is
convenient here to introduce the boson thermal mass
$$m_T=\sqrt{M_T^2+\lambda}\,.\eqnd\eSUSYbostherm $$
The parameter $m_T$ characterizes the decay of correlation
functions in space directions. Note that $M_T$, however, does not
characterize the decay of fermion correlations, because fermions
have no zero mode, the relevant parameter being
$\sqrt{M_T^2+\pi^2T^2}$.\par

Combining Eq.~\eNfivTtp~with the  $\theta =0$ part of the finite
temperature Eq.~\saddle{c} we obtain
$$ \eqalignno {\rho-\varphi^2/N &= \int {\d^2k\over (2\pi)^2 }
{1\over \omega_\varphi (k)}
 \left({1\over2} +{1\over \e^{  \omega_\varphi (k)/T}-1}\right)
 \cr &=\rho_c
  -{ T\over 2\pi}\ln\bigl(2\sinh(m_T/2T)\bigr),&\eqnd \esadSUSYNTa} $$
where  $\rho_c$ has been defined in Eq.\eRciiidef.

In the same manner, for fermions we introduce the function
$${\cal G}_2(M_T ,T) =  \int^\Lambda{\d^{d-1} k\over (2\pi)^{d-1}}{1\over \omega(k)
}\left({1\over2} -{1\over \e^{ \omega_\psi(k)/T}+1}\right)  \eqnd
\eNGNTtp  $$ with $\omega_\psi(k)=\sqrt{k^2+M_T^2}$. (This is the
fermion analogue of Eq.~\eNfivTtp\  in which one recognizes again
the sum of quantum and thermal contributions. Note that the
function ${\cal G}_2(M_T ,T)$ has a regular expansion in $M_T^2$
at $M_T=0$.) Combined with the $\bar\theta \theta $ part of
Eq.~\saddle{c}, it leads to
$$\eqalign{s-2F\varphi /N&=2M_T \int {\d^2k\over (2\pi)^2  } \left\{ {1\over 2 \omega _\varphi(k)}
\left({1\over 2}+{1\over \e^{ \omega_\varphi(k)/T}-1}\right)
\right. \cr&\quad \left.
 -{1\over 2 \omega _\psi(k)}\left({1\over2}
 -{1\over \e^{ \omega_\psi (k)/T}+1}\right)\right\}. }$$
Integrating we obtain
$$ s-2F\varphi /N =  {  T M_T\over  \pi}\left[
 \ln\bigl(2\cosh( M_T/ 2)\bigr)-\ln\bigl(2\sinh(m_T/2T)\bigr)\right].
\eqnd  \esadSUSYNTb $$

The action density ${\cal F}$ at finite temperature, to leading
order in $1/N$, is given by ($\beta =1/T$)
$${\cal F}={\cal S}_N/{V_2 \beta}\ , \eqnd\freeEnergyT$$
where ${\cal S}_N$ is given by Eq.~\eactSUSYNb~at constant
fields and $V_2$ is the two-dimensional volume.

The supertrace term in Eq.~\eactSUSYNb~ decomposes into
boson and fermion contributions. The boson  contribution
involves  ($\omega_n=2n{\pi T}$):
$$ \eqalignno{{1\over V_{d-1}}\tr\ln(-\nabla^2 + m_T^2) &=
\int {\d^{d-1}k\over {(2\pi)^{d-1}} }\sum_ {n\in {\cal Z}}
 \ln(\omega_n^2+ k^2+m_T^2) \cr &=2  \int
{\d^{d-1}k\over {(2\pi)^{d-1}} } \ln
\left[2\sinh\bigl({\beta\omega_\varphi(k)/2}\bigr)\right],&\eqnd\TRboson}$$
where $V_{d-1}$ is the $(d-1)$-dimensional volume. The result in
Eq.~\TRboson~ is given up to a mass independent infinite
constant.\par Similarly, for fermions (now $\omega_n=(2n+1){\pi
T}$)
$$\eqalignno{{1\over V_{d-1} \tr{\bf 1}}  \tr\ln(\sla{\partial} + M_T)  &=
  \ud  \int {\d^{d-1}k\over {(2\pi)^{d-1}} }\sum_
{n\in {\cal Z}}
 \ln(\omega_n^2+ k^2+M_T^2)  \cr
&= \int {\d^{d-1}k\over {(2\pi)^{d-1}} } \ln\left[2\cosh\bigl(
\omega_\psi (k)/  2T\bigr)\right].&
 \eqnd\eTRfermion\cr} $$

 Thus, one finds
$$\eqalignno{ {1\over V_2  } {\rm Str}\, &\ln\left [-\bar \D  \D +2L\right] =
{1\over V_2  }\tr\ln(-\partial^2 +M_T^2+\lambda) -{1\over V_2  }
\tr\ln(\sla{\partial}+M_T) \cr
  &=2  \int {\d^2k\over {(2\pi)^2} }
  \ln[2\sinh( \beta \omega _\varphi /2 ) ]
  -  2 \int {\d^2k\over {(2\pi)^2} }
  \ln[2\cosh( \beta  \omega _\psi/2)]    \cr
&={1 \over T}\rho_c (m_T^2-M_T^2) -{ 1\over 6\pi T}\left(
m_T^3-|M_T|^3\right) \cr &\quad +2\int {\d^2k\over  (2\pi)^2
}\left\{
  \ln[ 1-\e^{- \beta \omega _\varphi  } ]-  \ln[1+\e^{- \beta
  \omega _\psi}]\right\} .&\eqnd\SuperTraceT \cr} $$
The explicitly subtracted part reduces for $M_T=M$ to the $T= 0$
result. The other contributions to ${\cal F}$ are the same (up to
the change $M\mapsto M_T$) as in Eq.~\eSUSYV\ and, therefore, one
finds
$$\eqalignno{ {1\over N}{\cal F}&=
   -  {F^2\over2 N} +
M_T{F\varphi\over N} +   \lambda{\varphi^2\over 2N}  + \half s
(U'(\rho)-M_T)   -{1\over 12\pi}\left(  m_T^3 - |M_T|^3\right)\cr
&\quad + \half\lambda (\rho_c-\rho) +T\int {\d^2k\over  (2\pi)^2
}\left\{
  \ln[ 1-\e^{- \beta \omega _\varphi  } ]-
  \ln[1+\e^{- \beta  \omega _\psi}]\right\}  \,,
  &\eqnd\freeEnergyTiii \cr }$$
which reduces at $T=0$ to the action density of  Eq.~\eSUSYV,
${\cal F}(T=0)\equiv{\cal E}$.\par
Eq.~\freeEnergyTiii\ can be rewritten by using  Eq.~\esadSUSYNTa,
which is also obtained by setting to zero the $ \del/ \del\lambda
$ derivative of Eq.~\freeEnergyTiii, as well as $M_T-U'(\rho)=0$
from Eq.~\saddleComponB~(or equivalently, setting to zero the $
\del/ \del s $ derivative of Eq.~\freeEnergyTiii), one finds
$$\eqalignno{&{1\over N} {\cal F} = \half M_T^2 {\varphi^2\over N} + {1\over 24 \pi}(
m_T - |M_T|)^2(m_T +2|M_T|) \cr &\quad+{T\lambda\over 4\pi
}\ln(1-\e^{-m_T/T})+T \int {\d^2k\over {(2\pi)^2} } \{
\ln(1-\e^{-\beta \omega_\varphi} )
 - \ln(1 + \e^{-\beta \omega_\psi} ) \}.\hskip12mm
&\eqnd\FreeEnergy }$$ Eq.~\FreeEnergy~is the finite temperature
version of Eq.~\eSUSYground.
 \par
Inserting Eqs.~\esadSUSYNTa~and \esadSUSYNTb~(with $F=M_T\varphi$)
into  Eqs.~\saddleComponB~with $U( \rho )=\mu  \rho  +\half u \rho
^2$ one finds ($\mu_c=-u\rho_c$)
 $$\eqalignno{  M_T &=
  \mu-\mu_c   + u {\varphi^2\over N}
 -{u\over 2\pi}T\ln\left(2\sinh\left(\ud  m_T/T \right)\right) , &
 \eqnd\FermionMassi  \cr
{m_T^2-M_T^2 \over uM_T}&={2\varphi^2 \over N}     - {   T  \over
\pi}\left[ \ln\left(2\sinh\left(\ud  m_T/T
\right)\right)-\ln\left(2\cosh \left(\ud  |M_T|/T \right)\right)
\right], \hskip10mm & \eqnd  \esadSUSYNTla\cr} $$
 which are the   gap equations for  $T\neq 0$.
\medskip
{\it Solution to the saddle point equations:}
 ~We first note that the r.h.s.~of the gap Eq.~\FermionMassi\
 diverges when $m_T\to 0$.
The finite temperature system in three dimensions has the property
of a statistical system in two dimensions. In two dimensions a
spontaneous breaking of a continuous symmetry is impossible, due
to the  IR behaviour of a system with potential massless Goldstone
particles. Therefore, the $O(N)$ symmetry is never broken,
$\varphi=0$, and   thus, \eqna\eNTSUSYgapii
 $$\eqalignno{ { M_T \over T} &=
 { \mu-\mu_c\over T}     -{u\over 2\pi} \ln\left(2\sinh\left(\ud   m_T/T \right)\right) , &
 \eqnd\eNTSUSYgapii{a} \cr
{M_T^2-m_T^2\over uTM_T}&=    {   1  \over  \pi}\left[
\ln\left(2\sinh\left(\ud  m_T/T \right)\right)-\ln\left(2\cosh
\left(\ud  M_T /T \right)\right) \right]. \hskip2mm & \eqnd
\eNTSUSYgapii{b} \cr} $$
Note that while the expression of $\cal
F$ is complicated, its derivative with respect to $m_T^2 $ remains
simple:
$${1\over N}{\partial  {\cal F} \over \partial m_T^2 }={(m_T^2-M_T^2) \over 16\pi m_T  \tanh(m_T/2T)}.
\eqnn $$ Therefore, the minimum still occurs at $m_T=|M_T|$,  but
one verifies that $m_T=|M_T|$ is not a solution to the saddle
point equations. We have only found a lower bound on the free
energy density  $\cal F$ as a function of $M_T$:
$$ {\cal F}=NT \int {\d^2k\over (2\pi)^2 }\ln\tanh\left(\ud\beta \sqrt{M_T^2+k^2} \right).
\eqnd\eSUSYTFlow $$
Its derivative with respect to $M_T$ is
$${\partial {\cal F}\over \partial M_T}=-N{T M_T\over 2\pi}\ln\tanh(|M_T|/2T).$$
Therefore, the derivative has the sign of $M_T$. The lower bound
has a limit for $M_T\to 0$, which is thus an absolute lower bound:
$${\cal F}=-{7\over 8\pi}\zeta (3)NT^3\,.\eqnd\eSUSYFlow $$
Similarly, the  derivative of $\cal F$ with respect to $ M_T $ at
$m_T$ fixed is
$$\eqalignno{{\partial {\cal F}\over\partial  M_T }&=   { M_T\over 2\pi}
\left[
\ln\bigl(2\cosh(M_T/2T)\bigr)-\ln\bigl(2\sinh(m_T/2T)\bigr)\right]
\cr &={m_T^2-M_T^2 \over 2u}\,. \cr}$$
We now have to find the solutions to the saddle
 point equations and compare their free energies.\par
Note a first solution in the regime $T\to0$, $\mu<\mu_c$ with
$|m_T |\ll T$ and $|M_T\ll T$. Then, from \FermionMassi~we find the
boson thermal mass
$$m_T \sim T\e^{-2\pi(\mu_c-\mu)/uT}\,.$$
Eq.~\esadSUSYNTla~yields the fermion thermal mass
$$M_T \sim {m_T^2 \over \mu_c-\mu}\,.$$
Since both $m_T$ and $M_T$ are very small, the free energy is very
close to the lower bound \eSUSYFlow. The solution we have found
here is a precursor of the zero-temperature phase transition, and
corresponds to $\varphi\ne 0$ in region III and IV of
Fig.~\phases.
\par
Other solutions  exist in this regime but they converge, up to
exponential corrections, to the finite masses of the $T=0$
spectrum  and, thus, their free energy is much larger from the lower
bound \eSUSYTFlow.\par
 It is expected that even for $T$ larger and
$\mu\ge \mu_c$, the continuation of the small mass solution
remains the solution with lowest energy.\par In the following  we
take $m\equiv m_T$ as the free variable and, using
Eq.~\FermionMassi\ for $M_T=M_T(m,\varphi,\mu,u,T)$, we discuss
$$W(m,\varphi,T)={1\over N}{\cal F}
(m_T=m,M_T(m,\varphi,\mu,u,T),T).\eqnd\eWvar$$ Since, as described
above, at $T\neq 0$ there is no breaking of the $O(N)$ symmetry,
we will discuss mainly $\varphi=0$ in Eq.~\eWvar. For $T=0$,
Eq.~\eWvar\  results in $ {1\over N}{\cal F}(m,\varphi^2, T=0)
={1\over N}{\cal E}(m,\varphi^2)$ which has been analyzed above.
In particular we see the phase structure in Fig.\phases~ and the
interesting degeneracy found in Regions II and IV and exhibited in
Figs.\ \ZeroTone-\ZeroTtwo. Rather than changing the value of
$\mu-\mu_c$ and the coupling $u$ as has been done above at $T=0$,
we will be interested here to see the temperature dependence while
these parameters will be held fixed.

\par At finite temperature, the effective field theory describes
the interactions of the fermions and bosons with the heat bath.
This interaction acts like a source of soft breaking of
supersymmetry. The short distance behavior is cured at finite
temperature in a similar way it happens at $T=0$. We  now  show
that the general properties of the transitions as a function of
the temperature $T$ have a similar character as the transitions
seen at $T=0$ when the coupling constant is varied.

As seen in Fig.~\phases, at $T=0$ there are two regions in the $\{
\mu-\mu_c , u \}$ space where the vacua are degenerate.
These are (region {\bf II :} $\mu-\mu_c \geq 0$ and $u \geq u_c$
and region {\bf IV : } $\mu-\mu_c \leq 0$ and $u \leq u_c$).
Fig. \figTIVaa ~shows the ground state free
energy at $T=0$ in region {\bf IV}.

We discuss first region {\bf (II)}  $\mu-\mu_c \geq 0$
and $u \geq u_c$:

\noindent Here, as $T$ increases the ground state with the smaller
mass ($\ma\equiv m =m_+$) has a lower free energy than the heavier
one ($\ma \equiv m=m_- > m_+$) due to a sizable entropy
contribution.
(Fig.~\label{\figTIIa}).

  Peculiar transitions can occur in this system. If the
system was initially at $T=0$ ~in the ground state with a boson
mass $\ma=m_-$, it will eventually go, as $T$ is increased, into
the state with mass $\ma=m_+$, namely, into the lower mass ground
state. This is due to the entropy negative contribution to the
free energy forcing the system to favors the lowest mass state. On
the other hand, a system that started at a high temperature in the
state of low mass $\ma=m_+$ will stay in this state as the system
is cooled and will never roll back into the $m=m_-$ high mass
phase. Favoring the lowest  mass phase as the temperature
increases is a general effect that will occur in any physical
system that is initially (at $T=0$) mass degenerated. Here,
supersymmetry imply that the $m_+$ and $m_-$ vacua are at the same
energy $E=0$ at $T=0$.

\midinsert \epsfxsize=16.5cm \epsfysize=12cm \pspoints=2.pt \vskip
-5.8cm \hskip 1cm \epsfbox{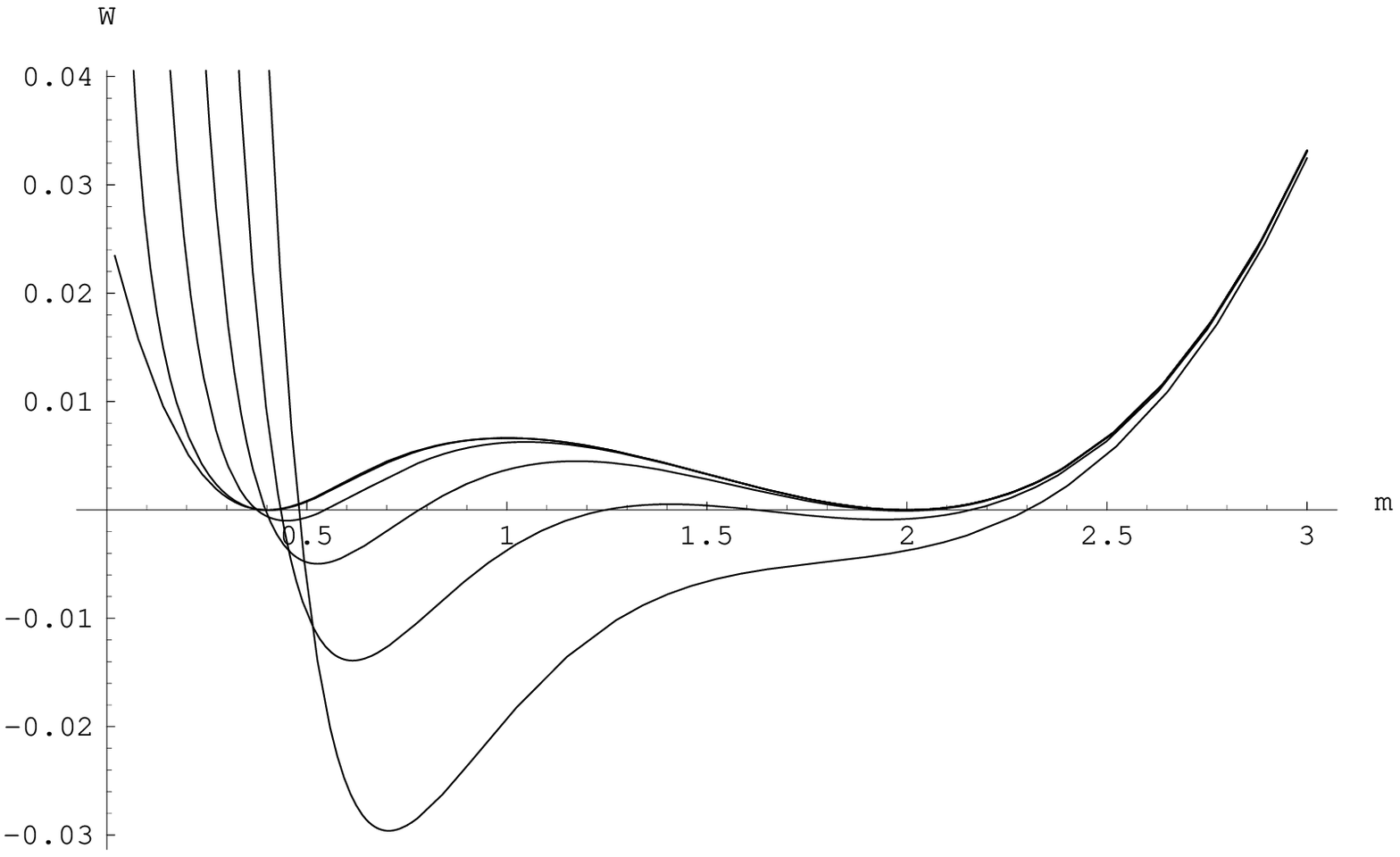} \figure{1mm}{The ground
state free energy ${1\over N} {\cal F}(m\equiv\ma,\varphi,T)$ as
given in Eq.~\FreeEnergy\ at $\varphi=0$ as a function of the
boson mass ($m$) at different temperatures. Here $\mu-\mu_c=1
$(this sets the mass scale) ,  $ u/u_c=1.5 $ and $T$ varies
between $T=0--0.5$. At $T = 0$ two degenerate $O(N)$ symmetric
phases exist  with a light $m=m_+$ and heavier $m=m_- > m_+$
massive boson (and fermion). As the temperature increases, the
light mass phase is stronger affected as its entropy increases
faster and becomes the only ground state.} \figlbl\figTIIa
\endinsert

Region {\bf (IV)}  $\mu \leq 0$ and $u \leq u_c$: here
one finds at $T=0$ two distinct degenerate phases. One is an
ordered phase ($\varphi \neq 0$) with a zero boson and fermion
mass, the other is a symmetric phase ($\varphi = 0$) with a
massive ($m=m_-$) boson and fermion  (as seen in
Fig.~\label{\figTIVaa}).

As the temperature increases, the ordered phase with the broken
$O(N)$ symmetry  ($m=0 , \varphi^2 \neq 0$) disappears into a
$\varphi^2 = 0$ symmetric phase and a very small mass ground state
$\ma=m\geq 0$ is created (as seen in Fig.~\label{\figTIVbb} and in
Fig.~\label{\figTIVa}).
 \midinsert \epsfxsize=17cm \epsfysize=14cm
\pspoints=2.pt \vskip -7.cm \hskip 1.5cm \epsfbox{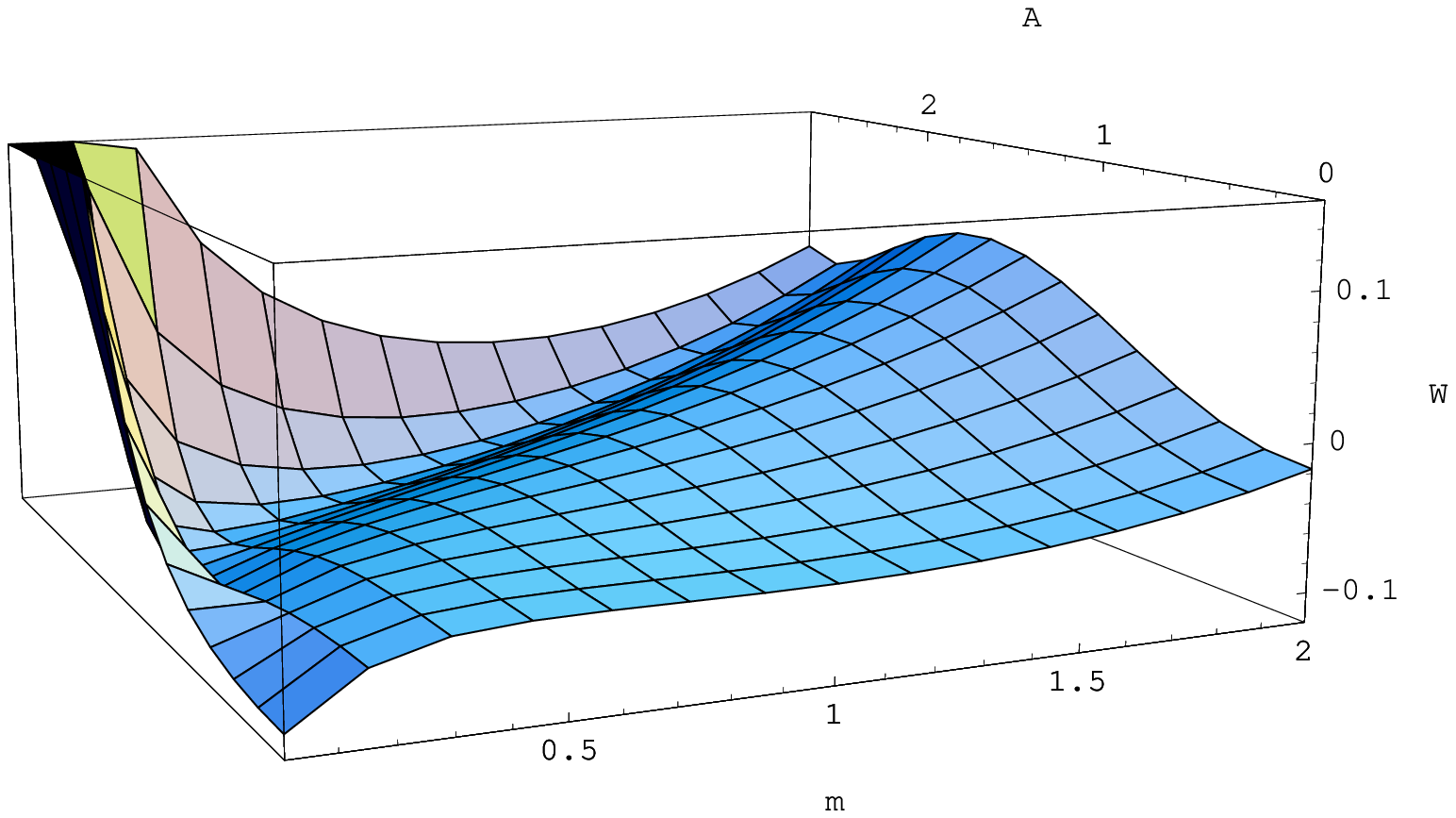}
\figure{1mm}{ Same as Fig.~\figTIVaa
 ~but the temperature has been increased from $T
= 0$ (in Fig.~\figTIVaa) to $T=0.7$ (here). ${1\over N} {\cal
F}(m\equiv\ma,\varphi,T)$ as given in
 Eq.~\FreeEnergy~ as a function of the
boson mass ($m$)  and $A$, where $A^2=\varphi^2/u_c$.
 Here $\mu-\mu_c=-1 $ and
 \ $ u/u_c=0.2 $. A non-degenerate $O(N)$ symmetric ground state
($\varphi = 0$) appears with a very small boson mass (the non-zero
mass is not seen here due to the limited resolution of the plot.
See Fig. \label{\figTIVa}~ ). } \figlbl\figTIVbb
\endinsert

Eventually, also the other $O(N)$ symmetric vacuum ($m=m_-,
\varphi^2 = 0$) goes  into the small mass $O(N)$-symmetric ground
state.
 \midinsert \epsfxsize=16cm \epsfysize=13cm \pspoints=2.pt
\vskip -6cm \hskip 1.5cm \epsfbox{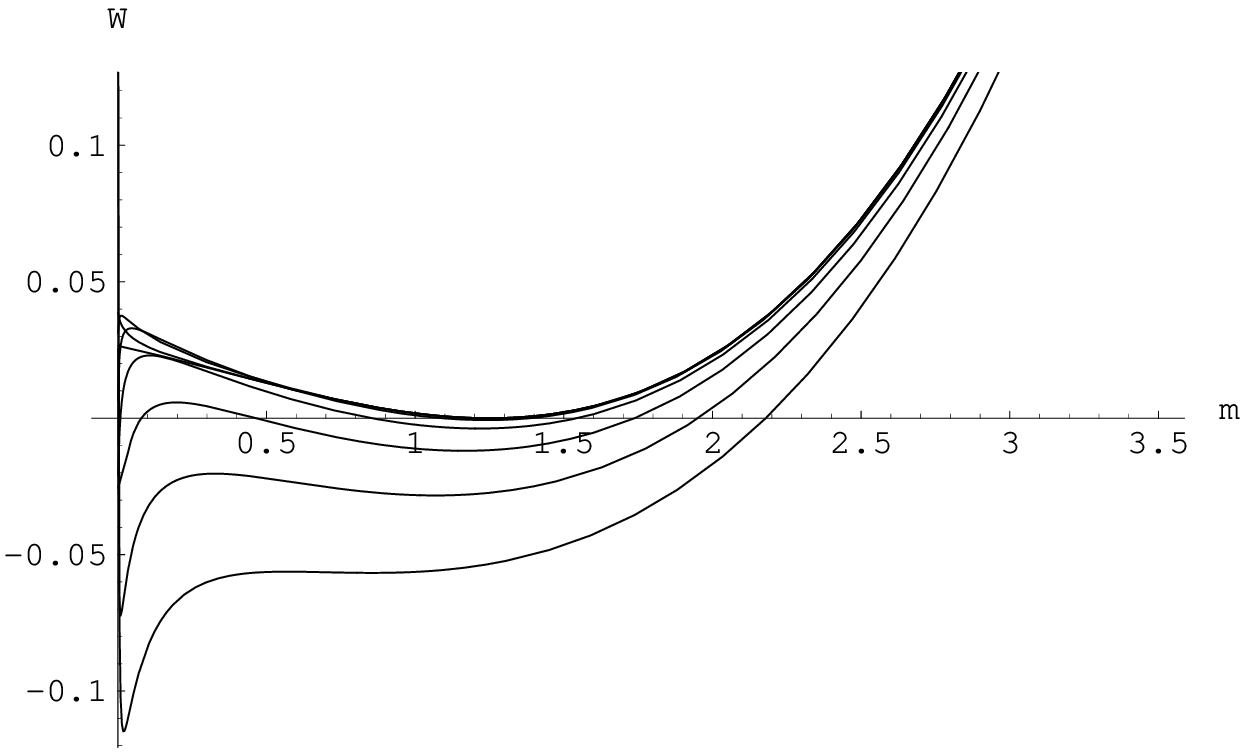} \figure{1mm}{This
figure displays the effect of increasing the temperature from $T =
0$ in Fig.~\figTIVaa~to $T$ that varies between $T=0 - 0.7$
(Fig.~\figTIVbb ~has $T=0.7$). The ground state free energy
${1\over N}{\cal F}(m\equiv\ma,\varphi,T)$ as given in Eq.
~\FreeEnergy at $\varphi=0$ is plotted as a function of the boson
mass ($m$)  at different temperatures. Here $\mu-\mu_c=-1 $, $
u/u_c=0.2 $. At $T = 0$ there are two degenerate phases: An
$O(N)$-symmetric phase, shown here, with a massive ($m=m_-$) boson
and fermion  and an ordered phase ($\varphi \neq 0$) with massless
particles (both phases are shown in Fig.~\figTIVaa). At finite
temperatures the $O(N)$ symmetry is restored (see Fig.~\figTIVbb )
and a small mass ground state appears, the heavy mass state decays
into the small mass ground state as seen here.} \figlbl\figTIVa
\endinsert

\noindent As in {\bf II}, a system that was initially, at $T=0$,
in the $O(N)$-symmetric phase ($\ma=m_- , \varphi^2 = 0$) will
decay into the smaller mass state when the temperature is raised. But upon
cooling the system in the small mass phase will roll into the
ordered phase ($\ma=m=0 , \varphi^2 \neq 0$) at $T=0$. The system
will never roll back into the symmetric phase ($\ma=m_- , \varphi
= 0$).

\section The supersymmetric $O(N)$ non-linear $\sigma $-model at finite temperature

The supersymmetric non-linear $\sigma $-model we consider here has
already been discussed above at zero temperature. The partition
function of the $O(N)$ supersymmetric non-linear $\sigma$-model in $d
$ dimensions is given by
$${\cal Z}=\int[\d\Phi][\d L] \e^{-{\cal S}(\Phi,L)} \eqnd\NLSMpartition$$
with
$${\cal S}(\Phi)={1\over 2 \kappa }\int\d^d x\,\d^2\theta\, \bar \D \Phi\cdot
\D \Phi +L(\Phi^2-N)  \,, \eqnd\eNLSMT $$ where $\kappa$ is a
constant. The scalar superfield $\Phi(x,\theta)$ in is an
$N$-component vector and the scalar superfield $L(x,\theta)$ is
parameterized as in the $T=0$ case.  After integrating out $N-1$
components of $\Phi$, leaving out $\phi=\Phi_1$, we obtain
$${\cal Z} = \int [d\phi][\d L] \e^{ - {\cal S}_N(\phi,L) }$$
where
$$   {\cal S}_N(\phi,L)= {1\over\kappa } \int\d^d x\,\d^2\theta  \left[ \half\bar \D \phi\cdot
\D \phi+  L \left(\phi^2- N \right) \right]    + {{N-1}\over 2}
{\rm Str}\, \ln( -{\bar \D }\D  +2 L )  .   \eqnd\eNLSTact   $$
Differentiating the  action  $     {\cal S}_N(\phi,L)$ with
respect to the superfields   $\phi$ and $L$,   one obtains
Eq.~\eSadNLSMa\ and the generalization of Eq.~\eSadNLSMb:
 \eqna\eSUSYNTnls
$$\eqalignno{ 2L\phi-\bar \D  \D \phi &=0 \,,&\eSUSYNTnls{a} \cr
 {N\over \kappa}-{\phi^2 \over N}&=\tr\Delta(k,\theta,\theta) .&\eSUSYNTnls{b} \cr}
 $$
The first equation is equivalent to $F=M_T\varphi$ and $\lambda
\varphi+FM_T=0$ while the second equation can be compared with
Eq.~\saddle{c}, and thus leads to (for $N\gg 1$ and $\psi=0$)
 \eqna\NLSMComponC
$$\eqalignno{ {1\over \kappa}-{\varphi^2\over N} &=\int {\d^dk\over (2\pi)^d }{1\over \omega_\varphi (k)}
 \left({1\over2} +{1\over \e^{  \omega_\varphi (k)/T}-1}\right), &\NLSMComponC{a}\cr
-{2F\varphi  \over N}&=2M_T \int {\d^dk\over (2\pi)^d  } \left\{
{1\over 2 \omega _\varphi(k)} \left({1\over 2}+{1\over \e^{
\omega_\varphi(k)/T}-1}\right) \right. \cr&\quad \left.
 -{1\over 2 \omega _\psi(k)}\left({1\over2} -{1\over \e^{ \omega_\psi (k)/T}+1}\right)\right\}  & \NLSMComponC{b}\cr} $$
with $\omega_\varphi (k)=\sqrt{m_T^2  +k^2}$ and  $\omega_ \psi
(k)=\sqrt{M_T^2  +k^2} $.
\medskip
{\it Dimension $d=3$.} As we have already discussed, the $d=3$
finite temperature theory is analogous from the point of phase
transitions to a two-dimensional theory. Therefore, the $O(N)$
symmetry remains unbroken and $\varphi=0$. We thus write the two
gap equations only  in this limit: \eqna\eNLSMsadTi
$$\eqalignno{
 {1\over \kappa _c}  - {1\over \kappa} &=
 {  T\over 2
\pi}\ln\bigl(2\sinh(m_T/2T)\bigr),&\eNLSMsadTi{a} \cr 0&= {  T
M_T\over 2 \pi}\left[\ln\bigl(2\sinh(m_T/2T)\bigr)-
\ln\bigl(2\cosh(M_T/2T)\bigr)\right]. &\eNLSMsadTi{b} } $$

The calculation of the ground state energy density to leading
order for $N\to\infty $ of the non-linear $\sigma $-model at
finite temperature follows steps similar to those of the
$(\Phi^2)^2$ field theory. The  free energy at finite temperature
is given by
$${\cal F}=T {\cal S}_N / V_2\,, \eqnd\NLSMFEnergyT$$
where ${\cal S}_N$ is given by Eq.~\eNLSTact~at constant fields
and $V_2$ is the two-dimensional volume. Here, the supertrace term
in Eq.~\eNLSTact~has to be replaced by the finite temperature form
as it appears in Eq.~\freeEnergyTiii. The free energy then is
given by
$$\eqalignno{{1\over N} {\cal F}&=      \ud(M_T^2-m_T^2)\left( {1\over\kappa } -{1\over\kappa_c}\right)
-{1\over12\pi}\left(m_T^3-|M_T|^3\right) \cr\quad & +T\int
{\d^2k\over  (2\pi)^2  }\left\{
  \ln[ 1-\e^{- \beta \omega _\varphi  } ]-  \ln[1+\e^{- \beta  \omega _\psi}]\right\}.
&\eqnd\NLSMFreeEnergyTiii \cr}$$
The  free energy in
Eq.~\NLSMFreeEnergyTiii~can be compared with Eq.~\eNLSMenergy\
which gives the ground state energy ${\cal F}$ at $T=0$.
\par
It can be verified that, after using the zero temperature limit of
Eq.~\eNLSMsadTi{a}, the energy density becomes
identical to the energy density of the $(\Phi^2)^2$
theory, up to a possible rescaling of the field.
It is now clear that the same mechanism works here.
Using Eq.~\eNLSMsadTi{a},
one indeed finds the expression \FreeEnergy\ (for $\varphi=0$):
$$\eqalignno{&{1\over N} {\cal F} =  {1\over 24 \pi}(
m_T - |M_T|)^2(m_T +2|M_T|) +{T \over 4\pi
}(m_T^2-M_T^2)\ln(1-\e^{-m_T/T})\cr &\quad+T \int {\d^2k\over
{(2\pi)^2} } \{ \ln(1-\e^{-\beta \omega_\varphi} )
 - \ln(1 + \e^{-\beta \omega_\psi} ) \}.\hskip12mm
&\eqnn\cr} $$
\medskip
{\it Solutions.} The derivative of $\cal F$ with respect to
$|M_T|$ at $m_T$ fixed is
$${\partial {\cal F}\over\partial |M_T|}={|M_T|T\over 2\pi}
\left[
\ln\bigl(2\cosh(M_T/2T)\bigr)-\ln\bigl(2\sinh(m_T/2T)\bigr)\right].$$
It is convenient to introduce the notation
$$X(\kappa ,T)=\exp\left[{2\pi \over T}\left({1\over \kappa_c}-{1\over \kappa }\right)\right].$$
Then, the behaviour of $\cal F$ depends on the position of $X$ with
respect to $2$:
$$X=2\ \Leftrightarrow\ T={2\pi\over \ln 2}\left({1\over \kappa_c}-{1\over \kappa }\right), $$
an equation which has a solution only for $\kappa >\kappa _c$.\par
 For $X<2$, the derivative vanishes at $M_T=0$ and is positive
for all $|M_T|>0$. Instead for $X>2$, the derivative vanishes both
at $M_T=0$ and
$$|M_T|=2T\ln\left[\ud(X+\sqrt{X^2-4})\right],\eqnd\eSUSYNTnlsMT $$
which are the two solutions of Eq.~\eNLSMsadTi{a}. The minimum of
$\cal F$ is located at the second value \eSUSYNTnlsMT. \par
We find, therefore, an interesting non-analytic behaviour:
$$\cases{M_T=0 &for $X<2$, \cr
M_T=2T\ln\left[\ud(X+\sqrt{X^2-4})\right] & for $X>2$ .\cr}$$
Eq.~\eNLSMsadTi{a} then yields directly the boson thermal mass
$$m_T=2T\ln\left[\ud(X+\sqrt{X^2+4})\right].$$
For $\kappa <\kappa _c$ and $T\to 0$, we find the asymptotic
behaviour
$$m_T\sim  TX(\kappa ,T),$$
which is exponentially small, and $M_T=0$.\par
For $\kappa >\kappa_c$ and $T\to 0$, both $m_T$ and $M_T$
converge toward the finite $T=0$ limit with exponential corrections.\par
 In the opposite high
temperature limit $T\gg | 1/\kappa -1/\kappa _c|$, we find that
$m_T$ is asymptotically proportional to $T$:
$$m_T\sim 2T\ln\bigl((1+\sqrt{5})/2\bigr),$$
while  $M_T=0$.\par
 It is not clear whether such a result can survive $1/N$ corrections.
 \medskip
{\it Dimension $d=2$.} In generic dimensions $2\le d\le 3$, a phase
transition is not even possible at finite temperature  and
$\varphi=0$ (in two dimensions it is even impossible at zero
temperature).  The gap equations take the form
$$\eqalignno{{1\over \kappa }&=\Omega _d(m_T)+T^{d-2}f_d(m_T^2/T^2),
&\eqnn \cr
0&=M_T\left[\Omega_d(|M_T|)-T^{d-2}g_d(M_T^2/T^2)-\Omega
_d(m)-T^{d-2}f_d(m_T^2/T^2)\right] , \hskip6mm &\eqnn \cr}$$
where $f_d$ and $g_d$ are defined by
$$\eqalignno{f_d(z)&=N_{d-1}\int_0^\infty {x^{d-2}\d x \over
\sqrt{x^2+z}}{1\over
\exp[\sqrt{x^2+z}]-1} \cr
&=N_{d-1}\int_{\sqrt{z}}^\infty(y^2-z)^{(d-3)/2}{\d y\over
\e^y-1} &\eqnd\eThermfz\cr} $$
and
$$\eqalignno{g_d(s)&= N_{d-1}\int_0^\infty {x^{d-2}\d x \over
\sqrt{x^2+s}}{1\over \exp[\sqrt{x^2+s}]+1}\cr
&=N_{d-1}\int_{\sqrt{s}}^\infty (y^2-s)^{(d-3)/2}{\d y\over \e^y
+1} .&\eqnd\eThermfiis\cr}$$
\par
For $d=2$, it is then convenient to introduce the physical mass $m$ solution of
$${1 / \kappa }=\Omega _2(m), $$
so that the equations become
$$\Omega _2(m)=\Omega _2(m_T)+  f_2(m_T^2/T^2) ,\eqnn $$
and either $M_T=0$ or
 $$ \Omega_2(|M_T|)- g_2(M_T^2/T^2)-
 \Omega _2(m_T)- f_2(m_T^2/T^2) =0\,.\eqnn $$
The first equation which determines $m_T$ is identical to the
equation one obtains  for the usual non-linear $\sigma $-model. At
low temperature $m_T=m$ up to exponentially small corrections. At
high temperature
$${T\over m_T}\sim{1\over \pi}\ln(m_T/m)\sim {1\over \pi}\ln( T/m),$$
a consequence of the domination of the zero mode and the UV
asymptotic freedom of the non-linear $\sigma $-model.\par
Combining both gap equations we find
$$ \Omega_2(|M_T|)- g_2(M_T^2/T^2)=\Omega _2(m).$$
\par
The analysis of the equation for $d=2$ shows that,
for $T$ large, the equation has no solution and thus
$M_T=0$, but is has a solution for $T$ small. The situation,
therefore, is similar to what has been encountered in three
dimensions. Again, an analysis of $LL$ propagator and $1/N$
corrections is required to understand whether this result survives
beyond the large $N$ limit.
\vfill\eject

\centerline{\bf Acknowledgements }

\noindent MM would like to thank W.A. Bardeen and J. Feinberg for
useful discussions. The warm hospitality of the Saclay and
Fermilab theory groups  is also appreciated.

\appendix {}

\vskip -1cm

\section Conventions: Supersymmetry and Majorana spinors in
$d=3$

 We briefly recall the properties of Majorana
spinors in three euclidean space dimensions and explain our
notation. In three dimensions the spin group is $SU(2)$. Then, a
spinor transforms like
$$\psi_U=U\psi\,, \quad U\in SU(2) \ \ .$$
The role of Dirac $\gamma$ matrices is played by the Pauli
$\sigma$ matrices, $\gamma_\mu\equiv \sigma_\mu$. Moreover,
$\sigma_2$ is antisymmetric while $\sigma_2\sigma_\mu$ is
symmetric. This implies
$$\sigma_2\sigma_\mu\sigma_2=-{}^T\!\sigma_\mu \ \Rightarrow\ U^*=\sigma_2 U
\sigma_2\,\ \ .$$ A Majorana spinor corresponds to a neutral
fermion and has only two independent components $\psi_1,\psi_2$.

The conjugated spinor is defined by (${}^T$ means transposed)
$$\bar\psi={}^T\!\psi \sigma_2 \ \Leftrightarrow
\ \bar\psi  _\alpha =i  \epsilon  _{\alpha \beta }\psi _\beta\,,\eqnd\eMajo $$
($\epsilon _{\alpha \beta }=-\epsilon _{\beta \alpha }$, $\epsilon
_{12}=1$) and thus $\bar\psi$ transforms like
$$[{}^T\!\psi \sigma_2]_U=[{}^T\!\psi \sigma_2] U^\dagger\,  .$$

One also defines a spinor of Grassmann coordinates:
$$\bar\theta={}^T\!\theta \sigma_2 \,\ \ ,$$
Since the only non-vanishing product is $\theta_1\theta_2$, we have
$$\bar\theta_\alpha\theta_\beta=\ud\delta_{\alpha\beta}
\bar\theta\cdot\theta\,\   .$$
The scalar product
$\bar\theta\cdot\theta$ is
$$\bar\theta\cdot\theta=-2i\theta_1\theta_2\ \Rightarrow
\theta_\alpha\bar\theta_\beta=i\delta_{\alpha\beta}\theta_1\theta_2
\,.$$
If $\bar \theta',\theta'$ is another pair of coordinates,
because $\sigma_2\sigma_\mu$ is symmetric, one finds
$$\bar\theta \sigma_\mu \theta'={}^T\!\theta\sigma_2\sigma_\mu\theta'=
-\bar\theta' \sigma_\mu \theta\,, \eqnd\esigtetii $$
and for the same reason
$$\bar \theta\psi=\bar\psi\theta\,.$$
Other useful identities are
$$(\bar\theta\psi)^2=-\ud (\bar\theta\theta)(\bar\psi\psi),\quad
(\bar\theta\sla{p}\psi)^2=\ud p^2(\bar\psi\psi)(\bar\theta\theta).
$$
It is convenient to integrate over $\theta_1,\theta_2$ with the
measure
$$\d^2\theta\equiv {i\over2}\d\theta_2\d\theta_1\, .$$
Then,
$$\int\d^2\theta\,\bar\theta_\alpha\theta_\beta=\ud
\delta_{\alpha\beta}\,,\quad
\int\d^2\theta\,\bar\theta\cdot\theta=1\,.$$
With this convention
the identity kernel $\delta^2(\theta'-\theta)$ in $\theta$ space
is
$$\delta^2(\theta'-\theta)=(\bar\theta'-\bar\theta)\cdot
(\theta'-\theta). \eqnd\eIdtheta $$
\medskip
{\it Superfields and covariant derivatives.} A superfield
$\Phi(\theta)$ has the expansion
$$\Phi(\theta)=\varphi+\bar\theta\psi+\ud\bar\theta
\theta F \,.\eqnd\PhiSupField$$
Again, although only two $\theta$
variables are independent, we define the covariant derivatives
$\D _\alpha$ and $\bar \D _\alpha$ ($\bar \D =\sigma_2 \D $):
$$\eqalign{\D _\alpha&\equiv
{\partial\over\partial\bar\theta_\alpha}-(\sla{\partial}
\theta)_\alpha \,,\cr \bar \D _\alpha&\equiv{\partial \over\partial
\theta_\alpha} -(\bar\theta\sla{\partial})_\alpha \,.\cr}$$ Then
the anticommutation relation is
$$\{\D _\alpha,\bar \D _\beta\}=-2 [\sla{\partial}]_{\alpha\beta}\,.$$
Also
$$\bar \D _\alpha \D _\alpha={\partial\over\partial\theta_\alpha}
{\partial\over\partial\bar\theta_\alpha}-(\bar\theta\sla{\partial})_\alpha
{\partial\over\partial\bar\theta_\alpha}
-{\partial\over\partial\theta_\alpha}(\sla{\partial}\theta)_\alpha
+\bar\theta_\alpha\theta_\alpha\partial^2\,.$$
Since the $\sigma_\mu$ are traceless, using the identity \esigtetii~one
verifies that $\bar \D  \D $ can also be written as
$$\bar \D _\alpha \D _\alpha={\partial\over\partial\theta_\alpha}
{\partial\over\partial\bar\theta_\alpha}-2(\bar\theta\sla{\partial})_\alpha
{\partial\over\partial\bar\theta_\alpha}
+\bar\theta_\alpha\theta_\alpha\partial^2\,,$$
and, therefore, in component form
$$\bar \D _\alpha \D _\alpha \Phi=2F-2\bar\theta\sla{\partial}\psi
 +\bar\theta\theta \partial^2\varphi\,.$$and
$$\left(\bar {\rm D}_\alpha {\rm D}_\alpha\right)^2=4\partial ^2.$$
\medskip

{\it Supersymmetry generators and WT identities.}
Supersymmetry is
generated by the operators
$$  Q_\alpha={\partial\over\partial\bar\theta_\alpha}+(\sla{\partial} \theta)_\alpha      \,,\quad \bar Q_\alpha ={\partial \over\partial \theta_\alpha}
+(\bar\theta\sla{\partial})_\alpha \,, $$ which anticommute with
${\rm D}_\alpha$ (and thus $\bar {\rm D}_\alpha$). Then,
$$\{\bar Q_\alpha, Q_\beta\}=2[\sla{\partial}]_{\alpha\beta}\,.$$
Supersymmetry implies WT identities for correlation functions. The
$n$-point function $ W^{(n)}(p_k,\theta_k)$ of Fourier components
satisfies
$$Q_\alpha W^{(n)}\equiv \left[\sum_k {\partial \over\bar \partial \theta^k_\alpha}
-i( \sla{p_k}\theta^k)_\alpha\right] W^{(n)}(p,\theta)=0\,.$$ To
solve this equation, we set
$$W^{(n)}(p,\theta)=F^{(n)}(p,\theta)\exp\left[-{i\over2n}\sum_{jk}\bar\theta_j
(\sla{p_j}-\sla{p_k})\theta_k\right], \eqnd\eSUSYWTs $$ where
$F^{(n)}$ is a symmetric function in the exchange
$\{p_i,\theta_i\} \leftrightarrow \{p_j,\theta_j\}$. It then
satisfies
$$\sum_k {\partial \over\partial \theta^k_\alpha}
F^{(n)}(p,\theta)=0\,,$$
that is, is translation invariant in $\theta$
space.\par
In the case of the two-point function, this leads to the
general form
$$\eqalignno{W^{(2)}(p,\theta',\theta)
&=A(p^2)\left[1+C(p^2)\delta^2(\theta'-\theta)\right]
\e^{i\bar\theta\slam{p}\theta'} .&\eqnd\eSUSYiipt \cr
&=A(p^2)\left[1+C(p^2) (\bar\theta'-\bar\theta) (\theta'-\theta)
+i\bar\theta\sla{p}\theta'-\frac{1}{4}p^2\bar\theta\theta\bar\theta'\theta'
\right] .\cr }$$
Since the vertex functions $\Gamma^{(n)}$ satisfy
the same WT identities, they take the same general form.

\vfill\eject

\listrefs

\vfill\eject

\bye